\documentclass[%
aip,
pof,
amsmath,amssymb,
reprint,%
]{revtex4-1}

\usepackage {CJK} % 中文名字
\usepackage{booktabs} %三线表
\usepackage{multirow} %三线表细线
\usepackage{graphicx}% Include figure files
\usepackage{dcolumn}% Align table columns on decimal point
\usepackage{bm}% bold math
\usepackage{caption}
\usepackage{subcaption}
\usepackage[utf8]{inputenc}
\usepackage[T1]{fontenc}
\usepackage{mathptmx}
\usepackage{etoolbox}
\usepackage{hyperref}
\makeatletter
\def\@email#1#2{%
	\endgroup
	\patchcmd{\titleblock@produce}
	{\frontmatter@RRAPformat}
	{\frontmatter@RRAPformat{\produce@RRAP{*#1\href{mailto:#2}{#2}}}\frontmatter@RRAPformat}
	{}{}
}%
\makeatother
\begin{document}
\begin{CJK*}{UTF8}{gbsn} % 中文名字做准备

\preprint{AIP/123-QED}

%\title[Sample title]{Sample Title:\\with Forced Linebreak}
\title{Two-relaxation-time regularized lattice Boltzmann model for Navier-Stokes equations}
% Force line breaks with \\
\author{Yuan Yu~({\CJKfamily{gbsn}余愿})}
\email{yuyuan@xtu.edu.cn}
\affiliation{ 
	School of Mathematics and Computational Science, Xiangtan University, Xiangtan, 411105, China
}%
\affiliation{ 
	National Center for Applied Mathematics in Hunan, Xiangtan, 411105, China
}%
\affiliation{ 
	Hunan Key Laboratory for Computation and Simulation in Science and Engineering, Xiangtan University, Xiangtan, 411105, China
}%

\author{Zuojian Qin~({\CJKfamily{gbsn}覃祚健})}%

\affiliation{ 
	School of Mathematics and Computational Science, Xiangtan University, Xiangtan, 411105, China
}%

\author{Siwei Chen~({\CJKfamily{gbsn}陈思伟})}
% \homepage{http://www.Second.institution.edu/~Charlie.Author.}
\affiliation{ 
	School of Mathematics and Computational Science, Xiangtan University, Xiangtan, 411105, China
}%

\author{Shi Shu~({\CJKfamily{gbsn}舒适})}
% \homepage{http://www.Second.institution.edu/~Charlie.Author.}
\affiliation{ 
	School of Mathematics and Computational Science, Xiangtan University, Xiangtan, 411105, China
}%

\author{Hai-zhuan Yuan~({\CJKfamily{gbsn}袁海专})}
% \homepage{http://www.Second.institution.edu/~Charlie.Author.}
\affiliation{ 
	School of Mathematics and Computational Science, Xiangtan University, Xiangtan, 411105, China
}%
\email{yhz@xtu.edu.cn}

\date{\today}% It is always \today, today,

\begin{abstract}
	In this paper, we develop a two-relaxation-time regularized lattice Boltzmann (TRT-RLB) model for simulating weakly compressible isothermal flows. A free relaxation parameter, $\tau_{s,2}$, is employed to relax the regularized non-equilibrium third-order terms. Chapman-Enskog analysis reveals that our model can accurately recover the Navier-Stokes equations (NSEs). Theoretical analysis of the Poiseuille flow problem demonstrates that the slip velocity magnitude in the proposed model is controlled by a magic parameter, which can be entirely eliminated under specific values, consistent with the classical TRT model. Our simulations of the double shear layer problem, Taylor-Green vortex flow, and force-driven Poiseuille flow confirm that the stability and accuracy of our model significantly surpass those of both the regularized lattice Boltzmann (RLB) and two-relaxation-time (TRT) models, even under super-high Reynolds numbers as $Re=10^7$. Concurrently, the TRT-RLB model exhibits superior performance in very high viscosity scenarios. The simulations of creeping flow around a square cylinder demonstrates the model's capability to accurately compute ultra-low Reynolds numbers as $Re=10^{-7}$. This study establishes the TRT-RLB model as a flexible and robust tool in computational fluid dynamics.
\end{abstract}

\maketitle
\end{CJK*}

\section{Introduction}

\quad In recent years, the lattice Boltzmann method (LBM) has gained significant attention in the realm of computational fluid dynamics (CFD) and related fields. Due to its unique features such as inherent parallelism, ease of handling complex boundary conditions, physical intuitiveness, and scalability, the LBM is increasingly applied in solving nonlinear partial differential
equations~\cite{chai2020multiple,wang2021modified,zhao2020block} and simulating complex fluid problems, including turbulent flow~\cite{yu2005dns,suga2015d3q27,jacob2018new,guo2020improved}, combustion~\cite{chen2008simple,lei2021study,tayyab2020hybrid}, multiphase interactions~\cite{li2016lattice,huang2015multiphase,yu2019versatile,montessori2018regularized}, evaporation and phase change~\cite{li2016lattice,li2016pinning,markus2011simulation,klassen2014evaporation}, fluid-structure interaction~\cite{li2019coupling,wang2023recent}, porous media flow~\cite{zhang2022near,wang2017simulation,zhao2018pore}, and chemical reactions~\cite{yu2002scalar,wang2012numerical}, among others. 
However, many challenging issues in the lattice Boltzmann method arise from the nonlinear collision term, such as numerical instability, lack of Galilean invariance in transport coefficients, and the fixed and unadjustable Prandtl number. Adopting non-uniform grids or more complex execution schemes, such as those in the series of works by Thomas Bellotti~\cite{bellotti2024initialisation,bellotti2023numerical}, is undoubtedly a promising approach. However, if we consider only the use of uniform standard lattices and employ the simplest collision and transport schemes, improving the implementation of the collision process may be a universally recognized effective method. The standard approach for handling non-linear collision terms is to employ the Bhatnagar-Gross-Krook (BGK) approximation~\cite{bhatnagar1954model}, known as the lattice BGK (LBGK) model, which was developed by Qian et~al.~\cite{qian1992lattice}. In this model, the probability particle distribution functions relax towards equilibrium at a single rate, a simplicity that has contributed to its widespread popularity. For problems close to equilibrium, the effectiveness of the BGK model can suffice.  However, it encounters stability issues at high Reynolds numbers and experiences diminished accuracy at low Reynolds numbers. 

In order to address the limitations of the standard lattice BGK model and enhance its robustness and effectiveness, several alternative collision models have been proposed. The multiple-relaxation-time (MRT) model~\cite{lallemand2000theory,d2002multiple,du2006multi} is performed in the moment space, enabling the independent control of relaxation rates for different moments, resulting in enhanced accuracy and stability. In the cascaded model~\cite{geier2006cascaded,ning2016numerical,premnath2009incorporating,fei2017consistent}, the relaxation is formulated in the central moment space, with these moments defined in the reference frame that moves with the local fluid motion. Numerical results~\cite{geier2006cascaded,ning2016numerical} demonstrate that the cascaded model exhibits superior stability compared to the MRT model. It is noteworthy that Geier et~al.~\cite{geier2015cumulant,geier2017parametrization} further refined their proposed cascaded model and introduced the so-called cumulant model. Both of these models can be viewed as the MRT-type model~\cite{shan2019central}, implying that they share some common drawbacks while improving model stability. The relaxation matrices in these models are controlled not only by the shear viscosity but also by the bulk viscosity and several free parameters, and a definitive approach to determining the optimal relaxation matrices is currently lacking. Moreover, these models incur substantial computational overhead. It is worth noting that the entropy lattice Boltzmann model proposed by Karlin and his coworkers~\cite{karlin1998maximum,chikatamarla2006entropy} demonstrates remarkable numerical stability. Nevertheless, its complexity and inefficiency arise from the need to solve iterative extremum problems at each grid in every time step, making its widespread application challenging. Fortunately, there are two collision models that combine the simplicity of the BGK model and  the accuracy and stability advantages of the MRT-type model: the two-relaxation-time (TRT) model~\cite{ginzburg2008study,ginzburg2008two} and the regularized model~\cite{latt2006lattice,mattila2017high,coreixas2017recursive}. The TRT model is characterized by two relaxation times, one of which is controlled by shear viscosity, while the other one remains free, allowing for adjustments to the model's accuracy and stability. D'Humi{\`e}res and Ginzburg~\cite{d2009viscosity} demonstrated that the numerical error becomes unaffected by fluid viscosity when both relaxation rates adhere to a constant relationship. Ginzburg et~al.~\cite{ginzburg2008study,Khirevich2015CoarseAF} further confirmed that the TRT model can eliminate the viscosity-dependent numerical slip caused by the bounce-back scheme at solid boundaries. Recently, Gsell et~al.~\cite{gsell2021lattice} proved that the TRT model is particularly well-suited for highly viscous flows.  By adjusting the effective Knudsen number and the viscous incompressibility factor, there is no longer a restriction on the shear viscosity relaxation time, as in the single-relaxation-time (SRT) model where $\tau_{s,1}<3$~\cite{gsell2021lattice}, and their numerical experiments indicate that the maximum relaxation time can reach up to $10^4$~\cite{gsell2021lattice}. They further revealed that the TRT model is particularly well-suited for simulating non-Newtonian fluids with locally variable viscosity. By appropriately controlling the local value of the viscous incompressibility factor, viscosity ratios as high as $10^{10}$ can be achieved. Another simple and effective model is the regularized lattice Boltzmann (RLB) model, which has its origins predating the initial proposal of the aforementioned improved collision models. As early as 1994, Ladd et~al.~\cite{ladd1994numerical} proposed the initial version of the regularized model, but unfortunately, it did not gain significant attention for quite a while, and both equilibrium and off-equilibrium functions were limited to second-order expansions. In this model, the relaxation process of the second-order off-equilibrium moments in this model includes two components: the relaxation of the traceless part of the second-order stress tensor controlled by the shear viscosity relaxation rate, and the relaxation of the trace part of the second-order stress tensor controlled by the bulk viscosity relaxation rate induced by discrete errors. Subsequently, Latt et~al.~\cite{latt2006lattice} proposed a more concise regularized model, in which the off-equilibrium second-order moments are no longer divided into traceless and trace parts but instead relax independently. Its straightforward formulation, requiring only an additional regularized step compared to the LBGK model, enhances its comprehensibility. Moreover, this model demonstrates superior robustness compared to LBGK, thus bringing regularized models into sharper focus among researchers. Following that, Zhang et~al.~\cite{zhang2006efficient} proposed higher-order regularized model based on Hermite expansion, which was utilized to simulate non-continuous flows at high Knudsen numbers. While the introduction of higher-order regularized models markedly improved stability and accuracy, it also heightened complexity and computational demands. Consequently, there was a noticeable lack of substantial advancements in the field for a considerable period afterward. In 2015, Malaspinas~\cite{malaspinas2015increasing} proposed an isothermal recursive regularized model, which efficiently reduced the complexity of higher-order models and simplified computations. Since then, the regularized-type models have gained increasingly successful applications. Coreixas et~al.'s recursive regularization model~\cite{coreixas2017recursive}, when applied to simulate the double shear layer problem, achieved remarkably high Reynolds numbers of up to $10^6$. The hybrid recursive regularized thermal lattice Boltzmann model proposed by Feng et~al.~\cite{feng2019hybrid,renard2021improved} enables simulations of subsonic and supersonic flows on standard lattice. Farag et~al.~\cite{farag2020pressure} proposed a pressure-based regularized model for simulating compressible flow problems, achieving a maximum Mach number of 1.5. The hybrid recursive regularized model proposed by Jacob et~al.~\cite{jacob2018new} demonstrates the outstanding performance in simulating large eddy simulation of weakly compressible flows. Moreover, the regularized model has achieved significant advancements across various application domains~\cite{tayyab2020hybrid,zhao2018pore}. Due to limitations in space, a comprehensive listing of these applications is not provided here. Several comparative or testing-oriented works have reaffirmed the significant advantages of the regularized model over other collision models. For example, Malaspinas~\cite{malaspinas2015increasing} demonstrated that the recursive regularized model exhibit superior numerical dissipation and dispersion properties compared to raw MRT~\cite{lallemand2000theory,d2002multiple,du2006multi} model. Kr{\"a}mer et~al.~\cite{kramer2019pseudoentropic} compared various collision models including MRT, entropic, quasi-equilibrium, regularized, and cumulant schemes, and they found that~\cite{kramer2019pseudoentropic} only the regularized model and the KBC model (the entropic MRT model proposed by Karlin, B{\"o}sch, and Chikatamarla~\cite{karlin2014gibbs}) were able to successfully reproduce the largest eddies on a $16 \times 16$ grid, while other conventional collision models had already failed. Compared to the KBC model~\cite{kramer2019pseudoentropic}, the regularized model more effectively suppressed spurious vortices~\cite{kramer2019pseudoentropic}.

Although the regularized model performs exceptionally well, it, like BGK model, suffers from viscosity-dependent discretization error. This error leads to two significant issues~\cite{gsell2021lattice}. On one hand, it results in numerical slip at the wall boundaries when implementing the bounce-back rule, severely limiting its applicability in simulating flow phenomena within confined geometries such as porous media and microchannels. On the other hand, it compromises the accuracy in simulating high-viscosity problems and fluids with rapidly changing viscosity, such as power-law fluids.

In order to further improve the numerical stability of the regularized model, mitigate viscosity-dependent discretization error, eliminate numerical slip, and enhance its precision in high-viscosity problems, this paper proposes a two-relaxation-time regularized lattice Boltzmann (TRT-RLB) model to simulate the isothermal weakly compressible flows. The rest of this paper is organized as follows. In Sec.~\ref{section2}, the TRT-RLB model proposed in this article is provided firstly. Secondly,  the Chapman-Enskog~(CE) analysis process for the recovery of macroscopic equations is provided. Lastly, the theoretical proof is given for the condition of eliminating numerical slip at Dirichlet boundary within this model. In Sec.~\ref{section3}, Four different benchmark problems, including double shear layer, decaying Taylor-Green vortex, Poiseuille flow, and creeping flow past a square cylinder, were simulated to investigate the performance of the proposed TRT-RLB model in terms of numerical accuracy, stability, convergence rate, the capability to eliminate numerical slip, and robustness in ultra-low Reynolds number  problems. Finally, some conclusions are given in Sec.~\ref{section4}.

\section{Methods}\label{section2}
\subsection{Governing equation}
\quad The $d$-dimensional Navier-Stokes equations (NSEs) with the forcing terms considered in this paper, utilizing Einstein's notation for simplicity, are given  as follows:
\begin{subequations}
	\label{eq1}
	\begin{equation} \label{eq1:1}
		\partial_t \rho +\partial_\alpha \left(\rho u_\alpha \right)=0,
	\end{equation}
	\begin{equation} \label{eq1:2}
	\partial_t \left(\rho u_\alpha \right)+\partial_\beta \left(\rho u_\alpha u_\beta+\rho c_s^{2}\delta_{\alpha \beta}\right)=\partial_\beta\mu\left(\partial_\alpha u_\beta+\partial_\beta u_\alpha\right)+F_\alpha.
	\end{equation}
\end{subequations}
where $\rho$ is the fluid density, $\mu$ is the dynamic viscosity, and $u_\alpha$ and $F_\alpha$ represent the components of fluid velocity and the body force in the $\alpha$ direction, respectively.

\subsection{Two-relaxation-time regularized lattice Boltzmann model}
\quad Inspired by the work of Zhao et al~.\cite{zhao2020block} in solving convection-diffusion equations, we employ a high-order Hermite expansion for off-equilibrium, retaining terms up to the third order. Furthermore, drawing inspiration from the work of Dellar et~al.~\cite{dellar2003incompressible} on orthogonal multiple-relaxation-time models, we apply independent relaxation times to relaxation processes of different orders.  These approaches have led us to propose a new off-equilibrium relaxation framework. The evolution equation of the new proposed TRT-RLB model for the NSEs given by Eqs.~(\ref{eq1}) can be written as

\begin{align}\label{eq2}
	f_i\left(x_\alpha+e_{i \alpha} \Delta t, t +\Delta t\right)=
	& f_i^{e q}\left(x_\alpha,t\right)+\left(1-\frac{1}{\tau_{s,1}}\right) w_i \frac{\mathcal{H}_{i, \alpha}}{c_s^2} \mathcal{A}_\alpha^{\mathrm{neq}}+\left(1-\frac{1}{\tau_{s,1}}\right) w_i \frac{\mathcal{H}_{i, \alpha \beta}}{2 c_s^4} \mathcal{A}_{\alpha \beta}^{\mathrm{neq}} \notag \\
	& +\left(1-\frac{1}{\tau_{s,2}}\right) w_i \frac{\mathcal{H}_{i, \alpha \beta \gamma}}{6 c_s^6} \mathcal{A}_{\alpha \beta \gamma}^{\mathrm{neq}}+G_i \Delta t+ \left(1-\frac{1}{2\tau_{s,1}}\right) F_i\Delta t,
\end{align}
where $f_{i}$ is the distribution function in the $i$-th direction of the discrete velocity $e_{i\alpha}$ with $i=0,...,(q-1)$ for a given D$d$Q$q$ lattice model, $\tau_{s,1}$ and $\tau_{s,2}$ are the dimensionless relaxation times, $f_i^{e q}$ is the equilibrium distribution function, $w_i$ is the weight coefficient, $c_s$ is the sound speed, and the definition of Hermite tensors up to the third order are
\begin{subequations}
	\label{eq3}
	\begin{equation} \label{eq3:1}
		\mathcal{H}_i=1,
	\end{equation}
	\begin{equation} \label{eq3:2}
		\mathcal{H}_{i, \alpha}=e_{i \alpha},
	\end{equation}
	\begin{equation} \label{eq3:3}
	\mathcal{H}_{i, \alpha \beta}=e_{i \alpha} e_{i \beta}-c_s^2 \delta_{\alpha \beta},
	\end{equation}
	\begin{equation} \label{eq3:4}
	\mathcal{H}_{i, \alpha \beta \gamma}=e_{i \alpha} e_{i \beta} e_{i \gamma}-c_s^2\left(e_{i \alpha} \delta_{\beta \gamma}+e_{i \beta} \delta_{\gamma \alpha}+e_{i \gamma} \delta_{\alpha \beta}\right),
	\end{equation}
\end{subequations}
where $\delta_{\alpha \beta}$ is the Kronecker delta. It should be noted that the following relationships~\cite{coreixas2017recursive} can be utilized to simplify the calculations:
\begin{subequations}
	\begin{equation}
		\mathcal{H}_{i, xxx}=\mathcal{H}_{i, yyy}\equiv0,
	\end{equation}
	\begin{equation}
		\mathcal{H}_{i, xxy}=\mathcal{H}_{i, xx}e_{i,y},
	\end{equation}
	\begin{equation}
		\mathcal{H}_{i, xyy}=\mathcal{H}_{i, yy}e_{i,x}.
	\end{equation}
\end{subequations}
In this work, the standard lattice model D$2$Q$9$ is adopted. Subsequently, the discrete velocity $e_{i\alpha}$ is determined as
\begin{gather*}
	\left(\begin{array}{c}e_{ix}\\
		e_{iy}\end{array}\right)=c\left[\begin{array}{ccccccccc}
		0 & 1 & 0 & -1 & 0 & 1 & -1 & -1 & 1 \\
		0 & 0 & 1 & 0 & -1 & 1 & 1 & -1 & -1
	\end{array}\right],
\end{gather*}
the weight coefficient $w_i$ is given by 
\begin{equation*}
	w_0=4/9,\quad w_{1,\cdots,4}=1/9,\quad w_{5,\cdots,8}=1/36,
\end{equation*}
and the sound speed is defined by $c_s=c/\sqrt{3}$ with $c=\Delta x/\Delta t$ being the particle speed, where $\Delta x$ and $\Delta t$ represent the spatial step and the time step, respectively. It is well known that, in the process of CE analysis, the discrete equilibrium third-order moments are required, and the difference between the discrete equilibrium third-order moments and the Maxwell equilibrium third-order moments directly leads to the infamous cubic velocity error term. To eliminate this error term, a straightforward approach is to expand the equilibrium state to third order while adding a compensatory term to the evolution equation~\cite{zhang2006efficient,shan2006kinetic,mattila2017high,feng2019hybrid}. From the perspective of recovering the macroscopic equations, due to the orthogonality of Hermite polynomials, expansions of fourth order and higher have no effect on the recovery of the equations. Although incorporating the fourth-order Hermite basis $\mathcal{H}_{i, xxyy}$ into the equilibrium expansion, as has been done in some works, is not complex, for simplicity's sake, we opt not to further increase the model's complexity for that negligible gain in precision. The third-order Hermite expansion form of equilibrium is given as
\begin{equation}\label{eq4}
	f_i^{e q}=w_i \rho \left\{\mathcal{H}_i+\frac{\mathcal{H}_{i, \alpha}}{c_s^2} u_\alpha+\frac{\mathcal{H}_{i, \alpha \beta}}{2 c_s^4} u_\alpha u_\beta+\frac{\mathcal{H}_{i, \alpha \beta \gamma}}{6 c_s^6} u_\alpha u_\beta u_\gamma\right\}.
\end{equation}
Once the expression for the off-equilibrium components is determined in Eq. (\ref{eq2}), the collision and streaming steps are subsequently defined. In this model, the off-equilibrium moments are presented in the following projection form~\cite{coreixas2017recursive}:
\begin{subequations}\label{eq9}
	\begin{equation}
		\mathcal{A}_\alpha^{\mathrm {neq }}=\sum_i \mathcal{H}_{i, \alpha}\left(f_i-f_i^{e q}\right),
	\end{equation}
	\begin{equation}
		\mathcal{A}_{\alpha \beta}^{\mathrm {neq }}=\sum_i \mathcal{H}_{i, \alpha \beta}\left(f_i-f_i^{e q}\right),
	\end{equation}
	\begin{equation}
		\mathcal{A}_{\alpha \beta \gamma}^{\mathrm {neq }}=\sum_i \mathcal{H}_{i, \alpha \beta \gamma}\left(f_i-f_i^{e q}\right),
	\end{equation}
\end{subequations}
where the equilibrium moments can be derived from macroscopic quantities to decrease computational load with
\begin{subequations}\label{eq9_2}
	\begin{equation}
		\sum_i \mathcal{H}_{i, \alpha}f_i^{e q}=\rho u_{\alpha},
	\end{equation}
	\begin{equation}
		\sum_i \mathcal{H}_{i, \alpha \beta}f_i^{e q}=\rho u_{\alpha} u_{\beta},
	\end{equation}
	\begin{equation}
		\sum_i \mathcal{H}_{i, \alpha \beta \gamma}f_i^{e q}=\rho u_{\alpha} u_{\beta} u_{\gamma},
	\end{equation}
\end{subequations}
To eliminate the cubic velocity error term, we can design a compensatory source term that matches the equilibrium distribution function Eq.~(\ref{eq4}). The specific proof process is referred to in the next section. Here, we directly present this newly proposed compensatory source term as follows:
\begin{equation}
	\label{eq_Gi}
	G_i=-w_i \frac{\mathcal{H}_{i, \alpha \beta}}{6 c_s^6}\left(1-\frac{1}{2\tau_{s,1}}\right)\partial_\gamma \Phi_{\alpha \beta \gamma},
\end{equation}
where
\begin{gather}
	\Phi_{\alpha \beta \gamma}=\left\{
	\begin{aligned}
		\rho &u_x^3,\quad &&\alpha=\beta=\gamma=x, \\
		\rho &u_y^3,\quad &&\alpha=\beta=\gamma=y, \\
		&0, &&\quad \text{otherwise},
	\end{aligned}
	\right. \\
	\partial_\gamma \Phi_{\alpha\beta\gamma}=\partial_x\left(\rho u_x^3\right)\delta_{\alpha x}\delta_{\beta x}+\partial_y\left(\rho u_y^3\right)\delta_{\alpha y}\delta_{\beta y}.
\end{gather}
$F_i$ is the forcing term, which is taken as~\cite{luo1998unified}
\begin{equation}
	\label{eq6} 
	F_i=w_i\left\{\frac{\mathcal{H}_{i, \alpha}}{c_s^2}F_\alpha+\frac{\mathcal{H}_{i, \alpha \beta}}{2 c_s^4}\left(F_\alpha u_\beta+u_\alpha F_\beta\right)\right\},
\end{equation}

In this study, based on the conservation laws of mass and momentum under collisions and the change of variables introduced by He et~al.\cite{heDiscreteBoltzmannEquation1998} to achieve second order accuracy in time with a body force, $\rho$ and $u_\alpha$ satisfy
\begin{equation}
	\sum_i f_i=\sum_i f_i^{e q}=\rho, 
\end{equation}
and
\begin{equation}
	\label{eq11} \sum_i e_{i\alpha}f_i=\rho u_\alpha-\frac{\Delta t}{2}F_\alpha.
\end{equation}

\subsection{The Chapman-Enskog analysis} \label{sec_CE}
The process of recovering macroscopic equations from the LBM through CE analysis is fully mature and well-understood. Here we not only need to demonstrate that the TRT regularized off-equilibrium relaxation model can accurately recover the macroscopic equations, but also need to prove that the auxiliary source term $G_i$ we proposed indeed eliminates the cubic velocity error terms. These aspects have not been addressed in previous research. To ensure the clarity and completeness of our proof, we will provide a detailed process. Thus, we now derive the NSEs given by Eqs.~(\ref{eq1}) and (\ref{eq2}) from the above-mentioned TRT-RLB model through the CE analysis. For this objective, the CE analysis is performed in the following multiscale expansions on the small parameter $\epsilon$:
\begin{equation}
	\begin{split}\label{eq12}
		f_i=f_i^{(0)}+&\varepsilon f_i^{(1)}+\varepsilon^2 f_i^{(2)}, \quad 
		\partial_t=\varepsilon \partial_t^{(1)}+\varepsilon^2 \partial_t^{(2)}, \quad 
		\partial_\alpha=\varepsilon \partial_\alpha^{(1)}, \\
		&G_i=\varepsilon G_i^{(1)}, \quad 
		F_i=\varepsilon F_i^{(1)}, \quad 
		F_\alpha=\varepsilon F_\alpha^{(1)}.
	\end{split}
\end{equation}
Utilizing Eqs.~(\ref{eq9}) and (\ref{eq12}), we have
\begin{subequations}\label{eq13}
	\begin{gather}
		\label{eq13a}
		\begin{align}
			\mathcal{A}_\alpha^{\mathrm {neq}}=\mathcal{A}_\alpha^{\mathrm {neq,(0)}}+\varepsilon \mathcal{A}_\alpha^{\mathrm {neq,(1)}}+\varepsilon^2 \mathcal{A}_\alpha^{\mathrm {neq,(2)}},
		\end{align}\\
		\label{eq13b}
		\begin{align}
			\mathcal{A}_{\alpha \beta}^{\mathrm{neq}}=\mathcal{A}_{\alpha \beta}^{\mathrm{neq,(0) }}+\varepsilon \mathcal{A}_{\alpha \beta}^{\mathrm {neq,(1)}}+\varepsilon^2 \mathcal{A}_{\alpha \beta}^{\mathrm{neq,(2)}},
		\end{align}\\
		\label{eq13c}
		\begin{align}
			\mathcal{A}_{\alpha \beta \gamma}^{\mathrm{neq}}=\mathcal{A}_{\alpha \beta \gamma}^{\mathrm{neq,(0) }}+\varepsilon \mathcal{A}_{\alpha \beta \gamma}^{\mathrm {neq,(1)}}+\varepsilon^2 \mathcal{A}_{\alpha \beta \gamma}^{\mathrm{neq,(2)}}.
		\end{align}
	\end{gather}
\end{subequations}
%where
%\begin{subequations}
%	\begin{gather}
	%		\begin{align}
		%			\mathcal{A}_\alpha^{\mathrm {neq,(0)}}=\sum_i \mathcal{H}_{i, \alpha}\left(f_i^{(0)}-f_i^{\mathrm{eq}}\right),\quad
		%			\mathcal{A}_\alpha^{\mathrm {neq,(1)}}=\sum_i \mathcal{H}_{i, \alpha}f_i^{(1)},\quad
		%			\mathcal{A}_\alpha^{\mathrm {neq,(2)}}=\sum_i \mathcal{H}_{i, \alpha}f_i^{(2)},\notag
		%		\end{align}\\
	%		\begin{align}
		%			\mathcal{A}_{\alpha \beta}^{\mathrm{neq,(0)}}=\sum_i \mathcal{H}_{i, \alpha \beta}\left(f_i^{(0)}-f_i^{\mathrm{eq}}\right),\quad
		%			\mathcal{A}_{\alpha \beta}^{\mathrm {neq,(1)}}=\sum_i \mathcal{H}_{i, \alpha \beta}f_i^{(1)},\quad
		%			\mathcal{A}_{\alpha \beta}^{\mathrm{neq,(2)}}=\sum_i \mathcal{H}_{i, \alpha \beta}f_i^{(2)},\notag
		%		\end{align}\\
	%		\begin{align}
		%			\mathcal{A}_{\alpha \beta \gamma}^{\mathrm{neq,(0) }}=\sum_i \mathcal{H}_{i, \alpha \beta \gamma}\left(f_i^{(0)}-f_i^{\mathrm{eq}}\right),\quad
		%			\mathcal{A}_{\alpha \beta \gamma}^{\mathrm {neq,(1)}}=\sum_i \mathcal{H}_{i, \alpha \beta \gamma}f_i^{(1)},\quad
		%			\mathcal{A}_{\alpha \beta \gamma}^{\mathrm{neq,(2)}}=\sum_i \mathcal{H}_{i, \alpha \beta \gamma}f_i^{(2)}.\notag
		%		\end{align}
	%	\end{gather}
%\end{subequations}
From Eq.~(\ref{eq4}) we can get
\begin{subequations}\label{sumfeq}
	\begin{gather}
		\label{sumfeqa}
		\begin{align}
			\sum_i e_{i\alpha}f_i^{\mathrm{eq}}=\rho u_\alpha,\quad
		\end{align}\\
		\label{sumfeqb}
		\begin{align}
			\sum_i e_{i\alpha}e_{i\beta}f_i^{\mathrm{eq}}=\rho u_\alpha u_\beta+\rho c_s^2\delta_{\alpha \beta},
		\end{align}\\
		\label{sumfeqc}
		\begin{align}
			\sum_i e_{i\alpha}e_{i\beta}e_{i\gamma}f_i^{\mathrm{eq}}=\rho u_\alpha u_\beta u_\gamma+\rho c_s^2 u_\zeta \Delta_{\alpha \beta \gamma \zeta}-\Phi_{\alpha \beta \gamma},
		\end{align}
	\end{gather}
\end{subequations}
where $\Delta_{\alpha \beta \gamma \zeta}=\delta_{\alpha \beta}\delta_{\gamma \zeta}+\delta_{\alpha \gamma}\delta_{\beta \zeta}+\delta_{\alpha \zeta}\delta_{\beta \gamma}$. Furthermore, the following results are obtained from Eqs.~(\ref{eq_Gi}) and (\ref{eq6}),
\begin{subequations}\label{sumGi}
	\begin{gather}
		\begin{align}
			\sum_i G_i=0,\quad 
			\sum_i e_{i\alpha}G_i=0,\quad
			\sum_i e_{i\alpha}e_{i\beta}G_i=-\left(1-\frac{1}{2\tau_{s,1}}\right)\partial_\gamma \Phi_{\alpha \beta \gamma},
		\end{align}\\
		\label{sumGia}
		\begin{align}
			\sum_i F_i=0,\quad 
			\sum_i e_{i\alpha}F_i=F_\alpha,\quad
			\sum_i e_{i\alpha}e_{i\beta}F_i=F_\alpha u_\beta + u_\alpha F_\beta.
		\end{align}
	\end{gather}
\end{subequations}
Based on Eqs.~(\ref{eq11}), (\ref{eq12}), (\ref{eq13a}) and (\ref{sumfeqa}), we obtain
\begin{gather}
	\mathcal{A}_\alpha^{\mathrm {neq,(0)}}=0, \quad \mathcal{A}_\alpha^{\mathrm {neq,(1)}}=-\frac{\Delta t}{2}F_\alpha^{(1)}, \quad 
	\mathcal{A}_\alpha^{\mathrm {neq,(2)}}=0.
\end{gather}
Applying the Taylor expansion to Eq.~(\ref{eq3}), we get
\begin{align}
	f_i&+\Delta t\left(\partial_t+e_{i \alpha} \partial_\alpha\right) f_i+\frac{\Delta t^2}{2}\left(\partial_t+e_{i \alpha} \partial_\alpha\right)^2 f_i
	= f_i^{e q}-\frac{\Delta t}{2}\left(1-\frac{1}{\tau_{s,1}}\right) w_i \frac{\mathcal{H}_{i, \alpha}}{c_s^2}F_\alpha \notag \\
	&+\left(1-\frac{1}{\tau_{s,1}}\right) w_i \frac{\mathcal{H}_{i, \alpha \beta}}{2 c_s^4} \mathcal{A}_{\alpha \beta}^{\mathrm{neq}}+\left(1-\frac{1}{\tau_{s,2}}\right) w_i \frac{\mathcal{H}_{i, \alpha \beta \gamma}}{6 c_s^6} \mathcal{A}_{\alpha \beta \gamma}^{\mathrm{neq}}+G_i \Delta t+\left(1-\frac{1}{2\tau_{s,1}}\right) F_i\Delta t.
	\label{eq15}
\end{align}
Substituting Eqs.~(\ref{eq12}) and (\ref{eq13}) into Eq.~(\ref{eq15}), one can get the $\varepsilon^0$, $\varepsilon^1$ and $\varepsilon^2$ scale equations as follows
\begin{subequations}
	\begin{gather}
		\label{eq16a}
		\begin{align}
			\varepsilon^{0}: \quad 
			f_i^{(0)}=f_i^{e q}+\left(1-\frac{1}{\tau_{s,1}}\right)w_i \frac{\mathcal{H}_{i, \alpha \beta}}{2 c_s^4} \mathcal{A}_{\alpha \beta}^{\mathrm{neq,(0)}}+\left(1-\frac{1}{\tau_{s,2}}\right)w_i \frac{\mathcal{H}_{i, \alpha \beta \gamma}}{6 c_s^6} \mathcal{A}_{\alpha \beta \gamma}^{\mathrm{neq,(0)}},
		\end{align}\\
		\label{eq16b}
		\begin{align}
			\varepsilon^{1}: \quad 
			f_i^{(1)}&+\Delta t \left(\partial_t^{(1)}+e_{i \alpha} \partial_\alpha^{(1)}\right) f_i^{(0)}=-\frac{\Delta t}{2}\left(1-\frac{1}{\tau_{s,1}}\right) w_i \frac{\mathcal{H}_{i, \alpha}}{c_s^2} F_\alpha^{(1)}+\left(1-\frac{1}{\tau_{s,1}}\right)w_i  \frac{\mathcal{H}_{i, \alpha \beta}}{2 c_s^4}\mathcal{A}_{\alpha \beta}^{\mathrm{neq,(1)}} \notag \\
			&+\left(1-\frac{1}{\tau_{s,2}}\right)w_i  \frac{\mathcal{H}_{i, \alpha \beta \gamma}}{6 c_s^6}\mathcal{A}_{\alpha \beta \gamma}^{\mathrm{neq,(1)}}+G_i^{(1)} \Delta t+\left(1-\frac{1}{2\tau_{s,1}}\right)F_i^{(1)} \Delta t,
		\end{align}\\
		\label{eq16c}
		\begin{align}
			\varepsilon^{2}: \quad 
			f_i^{(2)}&+\Delta t \partial_t^{(2)} f_i^{(0)} +\Delta t\left(\partial_t^{(1)}+e_{i \alpha} \partial_\alpha^{(1)}\right) f_i^{(1)}+\frac{\Delta t^2}{2}\left(\partial_t^{(1)}+e_{i \alpha} \partial_\alpha^{(1)}\right)^2 f_i^{(0)}\notag \\
			&=\left(1-\frac{1}{\tau_{s,1}}\right) w_i \frac{\mathcal{H}_{i, \alpha \beta}}{2 c_s^4} \mathcal{A}_{\alpha \beta}^{\mathrm{neq,(2) }}+\left(1-\frac{1}{\tau_{s,2}}\right) w_i \frac{\mathcal{H}_{i, \alpha \beta \gamma}}{6 c_s^6} \mathcal{A}_{\alpha \beta \gamma}^{\mathrm{neq,(2)}}.
		\end{align}
	\end{gather}
\end{subequations}
Multiplying Eq.~(\ref{eq16a}) by $e_{i\alpha}e_{i\beta}$ and summing it over $i$, and multiplying Eq.~(\ref{eq16a}) by $e_{i\alpha}e_{i\beta}e_{i\gamma}$ then summing it over $i$, we can obtain 
\begin{gather}
	\mathcal{A}_{\alpha \beta}^{\mathrm{neq,(0)}}=0, \quad 
	\mathcal{A}_{\alpha \beta \gamma}^{\mathrm{neq,(0)}}=0.
\end{gather}
Then, one can get
\begin{gather}
	f_i^{(0)}=f_i^{e q}.
\end{gather}
Summing Eq.~(\ref{eq16b}) over $i$, and multiplying Eq.~(\ref{eq16b}) by $e_{i\alpha}$ and $e_{i\alpha}e_{i\beta}$ then summing it over $i$ respectively, we have 
\begin{subequations}
	\begin{gather}
		\label{eq19a}
		\begin{align}
			\partial_t^{(1)} \rho+\partial_\alpha^{(1)}\left(\rho u_\alpha \right)=0,
		\end{align}\\
		\label{eq19b}
		\begin{align}
			\partial_t^{(1)}\left(\rho u_\alpha\right)+\partial_\beta^{(1)}\left(\rho u_\alpha u_\beta+\rho c_s^2\delta_{\alpha\beta}\right)=F_\alpha^{(1)},
		\end{align}\\
		\label{eq19c}
		\begin{align}
			\mathcal{A}_{\alpha \beta}^{\mathrm{neq,(1)}}=-\tau_{s,1}\Delta t & \left\{\partial_t^{(1)}\left(\rho u_\alpha u_\beta+\rho c_s^2 \delta_{\alpha \beta}\right)+\partial_\gamma^{(1)}\left(\rho u_\alpha u_\beta u_\gamma+\rho c_s^2 u_\zeta \Delta_{\alpha \beta \gamma \zeta}\right)-\frac{1}{2\tau_{s,1}}\partial_\gamma^{(1)}\Phi_{\alpha \beta \gamma}\right. \notag \\
			& \left.-\left(1-\frac{1}{2 \tau_{s,1}}\right)\left(F_\alpha^{(1)} u_\beta+ u_\alpha F_\beta^{(1)}\right)  \right\}.
		\end{align}
	\end{gather}
\end{subequations}
By multiplying both sides of Equation~(\ref{eq16b}) with $\frac{\Delta t}{2}(\partial_t^{(1)}+e_{i \alpha} \partial_\alpha^{(1)})$ and then substituting the obtained result into Equation~(\ref{eq16c}), followed by summing over $i$ and multiplying each term by $e_{i\alpha}$ before summing over $i$ again, we can derive
\begin{subequations}
	\begin{gather}
		\label{eq20a}
		\begin{align}
			\partial_t^{(2)} \rho=0,
		\end{align}\\
		\label{eq20b}
		\begin{align}
			\partial_t^{(2)}\left(\rho u_\alpha\right)&+\partial_\beta^{(1)}\left(1-\frac{1}{2\tau_{s,1}}\right)\mathcal{A}_{\alpha \beta}^{\mathrm{neq,(1)}}-\frac{\Delta t}{2}\partial_\beta^{(1)}\left(1-\frac{1}{2\tau_{s,1}}\right)\partial_\gamma^{(1)}\Phi_{\alpha \beta \gamma} \notag \\
			&+\frac{\Delta t}{2}\partial_\beta^{(1)}\left(1-\frac{1}{2\tau_{s,1}}\right)\left(F_\alpha^{(1)}u_\beta+u_\alpha F_\beta^{(1)}\right)=0 .
		\end{align}
	\end{gather}
\end{subequations}
Substituting Eq.~(\ref{eq19c}) into Eq.~(\ref{eq20b}), we can obtain
\begin{align}
	\label{eq21}
	\partial_t^{(2)} \left(\rho u_\alpha\right)-\Delta t \partial_\beta^{(1)}\left(\tau_{s,1}-\frac{1}{2}\right)&\left\{\partial_t^{(1)}\left(\rho u_\alpha u_\beta+\rho c_s^2\delta_{\alpha \beta}\right)+\partial_\gamma^{(1)}\left(\rho u_\alpha u_\beta u_\gamma+\rho c_s^2 u_\zeta \Delta_{\alpha \beta \gamma \zeta}\right)\right.\notag \\
	&-\left.\left(F_\alpha^{(1)}u_\beta+u_\alpha F_\beta^{(1)}\right)\right\}=0.
\end{align}
Based on Eqs.~(\ref{eq19a}) and (\ref{eq19b}) and after some algebraic manipulation, one can get
\begin{align}
	\label{eq22}
	\partial_t^{(1)}\left(\rho u_\alpha u_\beta+\rho c_s^2\delta_{\alpha \beta}\right)&+\partial_\gamma^{(1)}\left(\rho u_\alpha u_\beta u_\gamma+\rho c_s^2 u_\zeta \Delta_{\alpha \beta \gamma \zeta}\right)-\left(F_\alpha^{(1)}u_\beta+u_\alpha F_\beta^{(1)}\right)\notag \\
	=&\rho c_s^2\partial_\alpha^{(1)}u_\beta+\rho c_s^2 \partial_\beta^{(1)}u_\alpha,
\end{align}
then according to Eq.~(\ref{eq21}), we have
\begin{gather}
	\label{eq23}
	\partial_t^{(2)} \left(\rho u_\alpha\right)- \partial_\beta^{(1)}\left(\tau_{s,1}-\frac{1}{2}\right)\Delta t\rho c_s^2\left(\partial_\alpha^{(1)}u_\beta+\partial_\beta^{(1)}u_\alpha\right)=0.
\end{gather}
Taking Eq.~(\ref{eq19a})$\times \varepsilon+$Eq.~(\ref{eq20a})$\times \varepsilon^2$ and Eq.~(\ref{eq19b})$\times \varepsilon+$Eq.~(\ref{eq23})$\times \varepsilon^2$, we get
\begin{gather}
	\label{eq24}
	\partial_t \rho +\partial_\alpha \left(\rho u_\alpha \right)=0,\\
	\label{eq25}
	\partial_t \left(\rho u_\alpha \right)+\partial_\beta \left(\rho u_\alpha u_\beta+\rho c_s^{2}\delta_{\alpha \beta}\right)=\partial_\beta\mu\left(\partial_\alpha u_\beta+\partial_\beta u_\alpha\right)+F_\alpha,
\end{gather}
where the dynamic viscosity $\mu$ evidently needs to satisfy the following equation:
\begin{gather}
	\label{eq_mu}
	\mu=\rho c_s^2\left(\tau_{s,1}-\frac{1}{2}\right)\Delta t .
\end{gather}

In summary, we have demonstrated that the newly proposed TRT-RLB model, as mentioned before, can accurately recover to the macroscopic NSEs. Furthermore, it has been established that the relaxation time, $\tau_{s,2}$, remains unconstrained throughout the entire recovering process.
\subsection{Analysis of half-way bounce-back scheme for Dirichlet boundary conditions}
\quad In this section, we will use Poiseuille flow as an example to analyze the half-way bounce-back (HWBB) scheme for the Dirichlet boundary condition following the work of He et~al.~\cite{he1997analytic}. Our objective is to demonstrate the relationship between numerical slip and the relaxation times, $\tau_{s,1}$ and $\tau_{s,2}$. In  Poiseuille flow, we have~\cite{he1997analytic} 
\begin{gather}
	\label{eq_a1}
	\rho=const.,\quad \partial_x u_x=0,\quad \partial_x u_y=0,
\end{gather}
where $u_y=0$ and the body force acts solely in the $x$-direction, i.e., 
\begin{gather}
	\label{eq_a2}
	F_x=\rho g,\quad F_y=0,
\end{gather}
where $g$ denotes the acceleration due to the external forces acting on the fluids.
When the grid configuration of the standard HWBB scheme is adopted, the analytical solution of this problem can be derived as
\begin{gather}
	\label{eq_a3}
	u_x=\frac{4u_c}{N^2}y_i\left(N-y_i\right),
\end{gather}
where $y_i=i-0.5$ represents the location of $i$-th grid node in the vertical direction with $i=1,...,N$, where $N=L/\Delta x$ with $L$ being the channel width and the space step $\Delta x=1$, and $u_c=L^2 g/(8\nu)$ is the characteristic velocity with $\nu=\mu/\rho$ being the kinematic viscosity. The fulfillment of Eq.~(\ref{eq_a1}) implies that the compensating term $G_i$ is equal to zero. Thus, the evolution equation is comprised of the collision and streaming,
\begin{align}
	Collision: f_i^*\left(x_\alpha, t\right)=
	& f_i^{e q}\left(x_\alpha,t\right)+\left(1-\frac{1}{\tau_{s,1}}\right) w_i \frac{\mathcal{H}_{i, \alpha}}{c_s^2} \mathcal{A}_\alpha^{\mathrm{neq}}+\left(1-\frac{1}{\tau_{s,1}}\right) w_i \frac{\mathcal{H}_{i, \alpha \beta}}{2 c_s^4} \mathcal{A}_{\alpha \beta}^{\mathrm{neq}} \notag \\
	& +\left(1-\frac{1}{\tau_{s,2}}\right) w_i \frac{\mathcal{H}_{i, \alpha \beta \gamma}}{6 c_s^6} \mathcal{A}_{\alpha \beta \gamma}^{\mathrm{neq}}+ \left(1-\frac{1}{2\tau_{s,1}}\right) F_i\Delta t,
\end{align}
\begin{gather}
	Streaming: f_i\left(x_\alpha+e_{i \alpha} \Delta t, t +\Delta t\right)=f_i^*\left(x_\alpha, t\right),
\end{gather}
where $f_i^*$ denotes the post-collision distribution function. By substituting Eqs.~(\ref{eq4}), (\ref{eq9}), (\ref{eq6}) and (\ref{eq_a2}) into Eq.~(\ref{eq3}), then multiplying it by $e_{ix}$ and summing it over $i$, and based on Eq.~(\ref{eq_a1}) we can obtain the following equations, 
\begin{align}
	\label{eq_a4}
	c\left(f_1^j-f_3^j\right)=&c\left(f_1^j-f_3^j\right)-\frac{2}{3}\left(\frac{1}{\tau_{s,1}}-\frac{1}{\tau_{s,2}}\right)\left[c\left(f_5^j-f_6^j\right)+c\left(f_8^j-f_7^j\right)\right]\notag \\
	-&\frac{1}{3}\left(\frac{2}{\tau_{s,1}}+\frac{1}{\tau_{s,2}}\right)c\left(f_1^j-f_3^j\right) +\frac{\rho}{3\tau_{s,1}}\left[2u_{x,j}+\left(2\tau_{s,1}-1\right)g\Delta t\right]-\frac{\rho u_{x,j}u_{y,j}^2}{c^2\tau_{s,2}},
\end{align}
\begin{align}
	\label{eq_a5}
	c\left(f_5^j-f_6^j\right)=&c\left(f_5^{j-1}-f_6^{j-1}\right)-\frac{1}{6}\left\{\left(\frac{1}{\tau_{s,1}}-\frac{1}{\tau_{s,2}}\right)\left[c\left(f_1^{j-1}-f_3^{j-1}\right)-2c\left(f_8^{j-1}-f_7^{j-1}\right)\right]\right.\notag \\
	+&\left.2\left(\frac{2}{\tau_{s,1}}+\frac{1}{\tau_{s,2}}\right)c\left(f_5^{j-1}-f_6^{j-1}\right)\right\}+\frac{c+3u_{y,j-1}}{12c\tau_{s,1}}\rho\left[2 u_{x,j-1}+\left(2\tau_{s,1}-1\right) g\Delta t\right]\notag\\
	+&\frac{\rho u_{x,j-1}u_{y,j-1}^2}{2c^2\tau_{s,2}},
\end{align}
\begin{align}
	\label{eq_a6}
	c\left(f_8^j-f_7^j\right)=&c\left(f_8^{j+1}-f_7^{j+1}\right)-\frac{1}{6}\left\{\left(\frac{1}{\tau_{s,1}}-\frac{1}{\tau_{s,2}}\right)\left[c\left(f_1^{j+1}-f_3^{j+1}\right)-2c\left(f_5^{j+1}-f_6^{j+1}\right)\right]\right.\notag \\
	+&\left.2\left(\frac{2}{\tau_{s,1}}+\frac{1}{\tau_{s,2}}\right)c\left(f_8^{j+1}-f_7^{j+1}\right)\right\}+\frac{c-3u_{y,j+1}}{12c\tau_{s,1}}\rho\left[2 u_{x,j+1}+\left(2\tau_{s,1}-1\right)g\Delta t\right]\notag\\
	+&\frac{\rho u_{x,j+1}u_{y,j+1}^2}{2c^2\tau_{s,2}},
\end{align}
where $\left(u_{x,j},u_{y,j}\right)$ is the velocity vector and $f_i^j$ is the distribution function at $y=\left(j-0.5\right)\Delta x$. Eqs.~(\ref{eq_a4}), (\ref{eq_a5}) and (\ref{eq_a6}) are valid for $1\leq j\leq n$. 

Based on Eq.~(\ref{eq11}), we can rewrite Eq.~(\ref{eq_a4}) as
\begin{gather}
	\label{eq_a7}
	c\left(f_1^j-f_3^j\right)=\frac{2}{3}\rho u_{x,j}+\frac{2\tau_{s,2}-1}{3}\rho g\Delta t-\frac{\rho u_{x,j}u_{y,j}^2}{c^2},
\end{gather}
then we also obtain
\begin{gather}
	\label{eq_a8}
	c\left(f_5^j-f_6^j\right)+c\left(f_8^j-f_7^j\right)=\frac{1}{3}\rho u_{x,j}-\frac{4\tau_{s,2}+1}{6}\rho g\Delta t+\frac{\rho u_{x,j}u_{y,j}^2}{c^2}.
\end{gather}
Substituting Eq.~(\ref{eq_a7}) into Eqs.~(\ref{eq_a5}) and (\ref{eq_a6}), we have
\begin{align}
	c\left(f_5^j-f_6^j\right)=&\frac{1}{3}\left[\left(\frac{1}{\tau_{s,1}}-\frac{1}{\tau_{s,2}}\right)c\left(f_8^{j-1}-f_7^{j-1}\right)-\left(\frac{2}{\tau_{s,1}}+\frac{1}{\tau_{s,2}}-3\right)c\left(f_5^{j-1}-f_6^{j-1}\right)\right]\notag \\
	\label{eq_a9}
	&+\left(\frac{1}{\tau_{s,1}}+\frac{2}{\tau_{s,2}}\right)\frac{\rho u_{x,j-1}}{18}+\frac{u_{y,j-1}}{4c\tau_{s,1}}\left[2\rho u_{x,j-1}+\left(2\tau_{s,1}-1\right)\rho g \Delta t\right]\notag \\
	&-\frac{2\tau_{s,1}-10\tau_{s,1}\tau_{s,2}+\tau_{s,2}+4\tau_{s,2}^2}{36\tau_{s,1}\tau_{s,2}}\rho g\Delta t+\left(\frac{1}{\tau_{s,1}}+\frac{2}{\tau_{s,2}}\right)\frac{\rho u_{x,j-1}u_{y,j-1}^2}{6c^2},
\end{align}
\begin{align}
	c\left(f_8^j-f_7^j\right)=&\frac{1}{3}\left[\left(\frac{1}{\tau_{s,1}}-\frac{1}{\tau_{s,2}}\right)c\left(f_5^{j+1}-f_6^{j+1}\right)-\left(\frac{2}{\tau_{s,1}}+\frac{1}{\tau_{s,2}}-3\right)c\left(f_8^{j+1}-f_7^{j+1}\right)\right]\notag \\
	\label{eq_a10}
	&+\left(\frac{1}{\tau_{s,1}}+\frac{2}{\tau_{s,2}}\right)\frac{\rho u_{x,j+1}}{18}-\frac{u_{y,j+1}}{4c\tau_{s,1}}\left[2\rho u_{x,j+1}+\left(2\tau_{s,1}-1\right)\rho g \Delta t\right]\notag \\
	&-\frac{2\tau_{s,1}-10\tau_{s,1}\tau_{s,2}+\tau_{s,2}+4\tau_{s,2}^2}{36\tau_{s,1}\tau_{s,2}}\rho g\Delta t+\left(\frac{1}{\tau_{s,1}}+\frac{2}{\tau_{s,2}}\right)\frac{\rho u_{x,j+1}u_{y,j+1}^2}{6c^2}.
\end{align}
By adding the above two equations together, with the assistance of Eq.~(\ref{eq_a8}), one can obtain 
\begin{align}
	\label{eq_a11}
	c\left(f_5^j-f_6^j\right)&+c\left(f_8^j-f_7^j\right)=\left(\frac{1}{\tau_{s,1}}-1\right)\left[c\left(f_5^{j+1}-f_6^{j+1}\right)+c\left(f_8^{j-1}-f_7^{j-1}\right)\right]\notag \\
	&+\frac{2\tau_{s,1}-1}{6\tau_{s,1}}\rho\left(u_{x,j+1}+u_{x,j-1}\right)-\frac{2\tau_{s,1}-1}{4c\tau_{s,1}}\left(u_{y,j+1}-u_{y,j-1}\right)\rho g \Delta t \notag \\
	&-\frac{\rho}{2c\tau_{s,1}}\left(u_{x,j+1}u_{y,j+1}-u_{x,j-1}u_{y,j-1}\right)+\frac{1+4\tau_{s,1}+4\tau_{s,2}-8\tau_{s,1}\tau_{s,2}}{6\tau_{s,1}}\rho g \Delta t,\notag \\
	&+\left(\tau_{s,1}-\frac{1}{2}\right)\frac{\rho \left(u_{x,j+1}u_{y,j+1}^2+u_{x,j-1}u_{y,j-1}^2\right)}{\tau_{s,1}c^2}.
\end{align}
Based on Eqs.~(\ref{eq_a9}) and (\ref{eq_a10}), we can get
\begin{align}
	\label{eq_a12}
	c\left(f_5^{j+1}-f_6^{j+1}\right)=&\frac{1}{3}\left[\left(\frac{1}{\tau_{s,1}}-\frac{1}{\tau_{s,2}}\right)c\left(f_8^{j}-f_7^{j}\right)-\left(\frac{2}{\tau_{s,1}}+\frac{1}{\tau_{s,2}}-3\right)c\left(f_5^{j}-f_6^{j}\right)\right]\notag \\
	&+\left(\frac{1}{\tau_{s,1}}+\frac{2}{\tau_{s,2}}\right)\frac{\rho u_{x,j}}{18}+\frac{u_{y,j}}{4c\tau_{s,1}}\left[2\rho u_{x,j}+\left(2\tau_{s,1}-1\right)\rho g \Delta t\right]\notag \\
	&-\frac{2\tau_{s,1}-10\tau_{s,1}\tau_{s,2}+\tau_{s,2}+4\tau_{s,2}^2}{36\tau_{s,1}\tau_{s,2}}\rho g\Delta t+\left(\frac{1}{\tau_{s,1}}+\frac{2}{\tau_{s,2}}\right)\frac{\rho u_{x,j}u_{y,j}^2}{6c^2},
\end{align}
\begin{align}
	\label{eq_a13}
	c\left(f_8^{j-1}-f_7^{j-1}\right)=&\frac{1}{3}\left[\left(\frac{1}{\tau_{s,1}}-\frac{1}{\tau_{s,2}}\right)c\left(f_5^{j}-f_6^{j}\right)-\left(\frac{2}{\tau_{s,1}}+\frac{1}{\tau_{s,2}}-3\right)c\left(f_8^{j}-f_7^{j}\right)\right]\notag \\
	&+\left(\frac{1}{\tau_{s,1}}+\frac{2}{\tau_{s,2}}\right)\frac{\rho u_{x,j}}{18}-\frac{u_{y,j}}{4c\tau_{s,1}}\left[2\rho u_{x,j}+\left(2\tau_{s,1}-1\right)\rho g \Delta t\right]\notag \\
	&-\frac{2\tau_{s,1}-10\tau_{s,1}\tau_{s,2}+\tau_{s,2}+4\tau_{s,2}^2}{36\tau_{s,1}\tau_{s,2}}\rho g\Delta t+\left(\frac{1}{\tau_{s,1}}+\frac{2}{\tau_{s,2}}\right)\frac{\rho u_{x,j}u_{y,j}^2}{6c^2}.
\end{align}
Taking summation of Eqs.~(\ref{eq_a12}) and (\ref{eq_a13}), then substituting Eq.~(\ref{eq_a8}) into it, we can obtain
\begin{gather}
	\label{eq_a14}
	c\left(f_5^{j+1}-f_6^{j+1}\right)+c\left(f_8^{j-1}-f_7^{j-1}\right)=\frac{1}{3}\rho u_{x,j}+\frac{5-4\tau_{s,2}}{6}\rho g\Delta t+\frac{\rho u_{x,j}u_{y,j}^2}{c^2}.
\end{gather}
Substituting Eqs.~(\ref{eq_a8}) and (\ref{eq_a14}) into Eq.~(\ref{eq_a11}), we have
\begin{align}
	\label{eq_a15}
	\frac{u_{x,j+1}u_{y,j+1}-u_{x,j-1}u_{y,j-1}}{2c}=&\frac{2\tau_{s,1}-1}{6}\left(u_{x,j+1}-2u_{x,j}+u_{x,j-1}\right)-\left(\tau_{s,1}-\frac{1}{2}\right)\frac{u_{y,j+1}-u_{y,j-1}}{2c}g \Delta t \notag \\
	+&\left(\tau_{s,1}-\frac{1}{2}\right)\frac{u_{x,j+1}u_{y,j+1}^2-2u_{x,j}u_{y,j}^2+u_{x,j-1}u_{y,j-1}^2}{c^2}+g\Delta t.
\end{align}
On the other hand, from $\rho u_{y,j}=c\left(f_2^j+f_5^j+f_6^j-f_4^j-f_7^j-f_8^j\right)$ and $\rho=\sum_i f_i^j$, we can easy to prove the following equations,
\begin{gather}
	\label{eq_a16}
	u_{y,j+1}^2-u_{y,j-1}^2=\left(2\tau_{s,1}-1\right)\left(u_{y,j+1}-2u_{y,j}+u_{y,j-1}\right)c, \\
	\label{eq_a17}
	u_{y,j+1}^2-2u_{y,j}^2+u_{y,j-1}^2=\left(u_{y,j+1}-u_{y,j-1}\right)c.
\end{gather}
From Eqs.~(\ref{eq_a16}) and (\ref{eq_a17}), we can obtain
\begin{gather}
	\label{eq_a18}
	u_{y,j}=u_y=const.
\end{gather}
Based on Eqs.~(\ref{eq_a15}) and (\ref{eq_a18}), and with the aid of the relationship between the kinematic viscosity $\nu$ and the relaxation time $\tau_{s,1}$, one can get the following equivalent difference equation of present model as 
\begin{align}
	\label{eq_a19}
	\frac{u_{x,j+1}u_y-u_{x,j-1}u_y}{2\Delta x}=\left(1+3 u_y^2 \frac{\Delta t^2}{\Delta x^2}\right)\nu\frac{u_{x,j+1}-2u_{x,j}+u_{x,j-1}}{\Delta x^2}+g.
\end{align}
To determine the slip velocity, an analysis of the bottom nodes $\left(j=1\right)$ is necessary (the analysis of the top nodes is the same). Based on the HWBB scheme, and with the help of the $u_y=0$, we have
\begin{gather}
	\label{eq_a20}
	f_2^1=f_4^{1,*},\quad f_5^1=f_7^{1,*},\quad f_6^1=f_8^{1,*}.
\end{gather}
Based on the above equations, we can obtain
\begin{gather}
	\label{eq_a21}
	c\left(f_5^1-f_6^1\right)=-c\left(f_8^{1,*}-f_7^{1,*}\right).
\end{gather}
Substituting Eqs.~(\ref{eq_a8}) and (\ref{eq_a13}) into Eq.~(\ref{eq_a21}), one can get
\begin{gather}
	\label{eq_a22}
	c\left(f_8^1-f_7^1\right)=\frac{2\tau_{s,1}+1}{6}\rho u_{x,1}-\frac{8\tau_{s,1}\tau_{s,2}-4\tau_{s,1}+4\tau_{s,2}-1}{12}\rho g \Delta t.
\end{gather}
Substituting Eq.~(\ref{eq_a8}) into Eq.~(\ref{eq_a10}), we have
\begin{gather}
	\label{eq_a23}
	c\left(f_8^1-f_7^1\right)=\frac{1-\tau_{s,1}}{\tau_{s,1}}c\left(f_5^2-f_6^2\right)+\frac{2\tau_{s,1}-1}{6\tau_{s,1}}\rho u_{x,2}-\frac{8\tau_{s,1}\tau_{s,2}-4\tau_{s,1}-4\tau_{s,2}-1}{12\tau_{s,1}}\rho g \Delta t.
\end{gather}
Substituting Eq.~(\ref{eq_a12}) into Eq.~(\ref{eq_a23}) with the aid of Eq.~(\ref{eq_a8}), one can obtain that 
\begin{gather}
	\label{eq_a24}
	c\left(f_8^1-f_7^1\right)=\frac{\rho}{6}\left[\left(1-\tau_{s,1}\right)u_{x,1}+\tau_{s,1}u_{x,2}\right]-\frac{8\tau_{s,1}\tau_{s,2}-4\tau_{s,1}-4\tau_{s,2}-1}{12\left(2\tau_{s,1}-1\right)}\rho g \Delta t,
\end{gather}
then substituting Eq.~(\ref{eq_a24}) into Eq.~(\ref{eq_a22}) yields
\begin{gather}
	\label{eq_a25}
	3u_{x,1}-u_{x,2}=\frac{8\tau_{s,1}\tau_{s,2}-4\tau_{s,1}-4\tau_{s,2}+5}{2\tau_{s,1}-1}g\Delta t.
\end{gather}
From Eq.~(\ref{eq_a3}), we can get
\begin{gather}
	\label{eq_a26}
	u_{x,1}=\frac{4u_c}{N^2}\frac{1}{2}\left(N-\frac{1}{2}\right)+u_s,\\
	\label{eq_a27}
	u_{x,2}=\frac{4u_c}{N^2}\frac{3}{2}\left(N-\frac{3}{2}\right)+u_s,
\end{gather}
where $u_s$ is the slip velocity.

By substituting Eqs.~(\ref{eq_a26}) and (\ref{eq_a27}) into Eq.~(\ref{eq_a25}), one can derive the numerical slip, 
\begin{gather}
	\label{eq_a28}
	u_s=\frac{16\tau_{s,1}\tau_{s,2}-8\tau_{s,1}-8\tau_{s,2}+1}{4\left(2\tau_{s,1}-1\right)}g\Delta t.
\end{gather}
Let $u_s=0$, we can obtain $\Lambda_s=3/16$ with the magic parameter $\Lambda_s$ being defined by
\begin{gather}
	\label{eq_a29}
	\Lambda_s=\left(\tau_{s,1}-\frac{1}{2}\right)\left(\tau_{s,2}-\frac{1}{2}\right).
\end{gather}

In summary, we demonstrate that the numerical slip of our proposed TRT-RLB model is related to the magic parameter $\Lambda_s$, and it can be completely eliminated for the HWBB scheme when $\Lambda_s=3/16$. This result is consistent with the conclusions drawn in the TRT model.

\section{Numerical results}\label{section3}
\quad In this section, we first conduct the simulation of the double shear layer flow to evaluate the stability. Then the 2D decaying Taylor-Green vortex flow is simulated to test grid convergence and accuracy. Furthermore, to validate the expression [Eq.~(\ref{eq_a29})] numerically, the force-driven Poiseuille flow is performed. Finally, the creeping flow past a square cylinder is simulated to verify the robustness of the model in ultra-low Reynolds number problems. In these simulations, for error measurement the following relative $L_2$ error is used,
\begin{gather}\label{eq_n1}
	L_2=\sqrt{\frac{\sum_{\mathbf{x}}\left(q_{n}\left(\mathbf{x},t\right)-q_{a}\left(\mathbf{x},t\right)\right)^2}{\sum_{\mathbf{x}}q_{a}^2\left(\mathbf{x},t\right)} },
\end{gather}
where $q_a\left(\mathbf{x},t\right)$ and $q_n\left(\mathbf{x},t\right)$ represent the analytical and numerical solutions, respectively. In addition, the following convergence criterion is adopted for steady flows:
\begin{gather}\label{eq_n2}
	\max(\left|\frac{u_\alpha\left(\mathbf{x},t+10000\Delta t\right)-u_\alpha\left(\mathbf{x},t\right)}{u_c}\right|)<10^{-7}.
\end{gather}

\subsection{Double shear layer}
\quad We first consider the double shear layer problem, that is widely recognized as a great option to assess the stability of numerical methods~\cite{coreixas2017recursive,dellar2014lattice,mattila2015investigation,mattila2017high}. Let us consider a two-dimensional periodic domain with $\left(x,y\right)\in \left[0,L\right]^2$. The initial conditions consist of two longitudinal shear layers and a transverse perturbation, i.e.,
\begin{gather}
	u_x\left(\mathbf{x},t=0\right)=\left\{
	\begin{aligned}
		u_c \tanh\left[\kappa\left(\frac{y}{L}-\frac{1}{4}\right)\right],\quad \frac{y}{L}\leq\frac{1}{2},\\
		u_c \tanh\left[\kappa\left(\frac{3}{4}-\frac{y}{L}\right)\right],\quad \frac{y}{L}>\frac{1}{2},
	\end{aligned}
	\right.
\end{gather}
and
\begin{gather}
	u_y\left(\mathbf{x},t=0\right)=u_c \delta \sin\left[2\pi\left(\frac{x}{L}+\frac{1}{4}\right)\right],
\end{gather}
where $\kappa$ controls the thickness of the shear layer, and $\delta$ is a small parameter controlling initial perturbation. In simulations, we set $\left(\kappa,\delta\right)=\left(80,0.05\right)$ and the particle speed $c=1$ with $\Delta x=L/N$, where $N=128$. The Reynolds and Mach numbers are $Re=\rho_0 u_c L/\mu$ and $Ma=u_c/c_s$, respectively, with $L=1$. In order to quantify the impact of the introduced relaxation time $\tau_{s,2}$ on stability, we investigated the maximum computable critical Mach number $Ma_{c}$ with $\Delta Ma=0.01$ that ensures stable simulation up to $t/t_c=2$ for various parameter values of $1/\tau_{s,2}$ ranging from $0.1$ to $2$ at different Reynolds numbers, where $t_c=L/\left(c_s Ma \right)$. Three common collision models, including BGK model, TRT model, and RLB model~\cite{coreixas2017recursive}, were employed for comparison with the proposed TRT-RLB model in this study. Specifically, the off-equilibrium in the RLB model has been expanded to third order, akin to the present model, rather than adopting the simplest RLB model proposed by Latt et~al.~\cite{latt2006lattice}. In other words, when $\tau_{s,1}=\tau_{s,2}$, the TRT-RLB model degenerates into the RLB model.

\begin{figure}[htbp]
	\centering
	
	\begin{subfigure}{0.32\textwidth}
		\includegraphics[width=\textwidth]{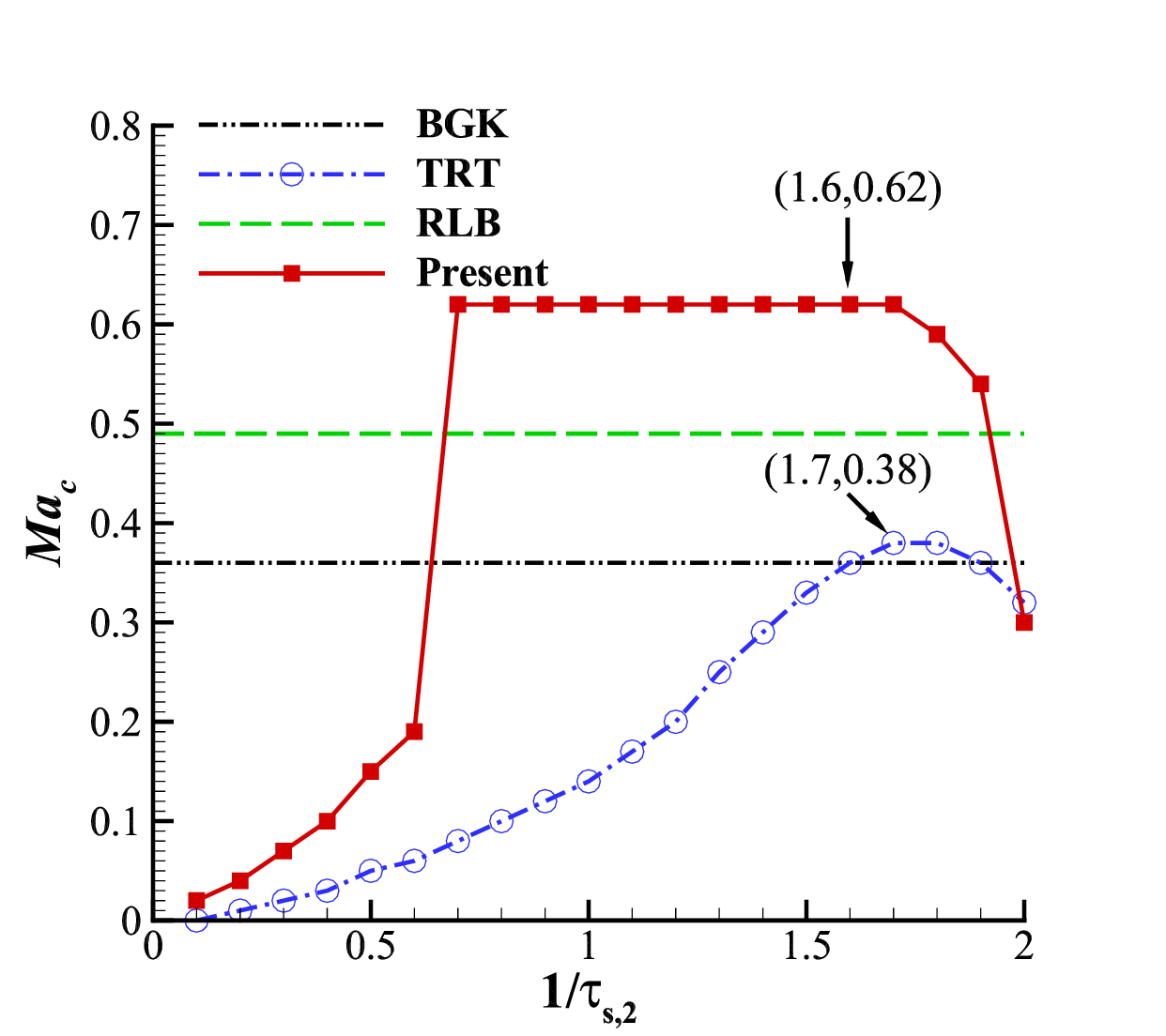}
		\caption{$Re=5\times 10^3$}
		\label{fig:sub1}
	\end{subfigure}
	\hfill
	\begin{subfigure}{0.32\textwidth}
		\includegraphics[width=\textwidth]{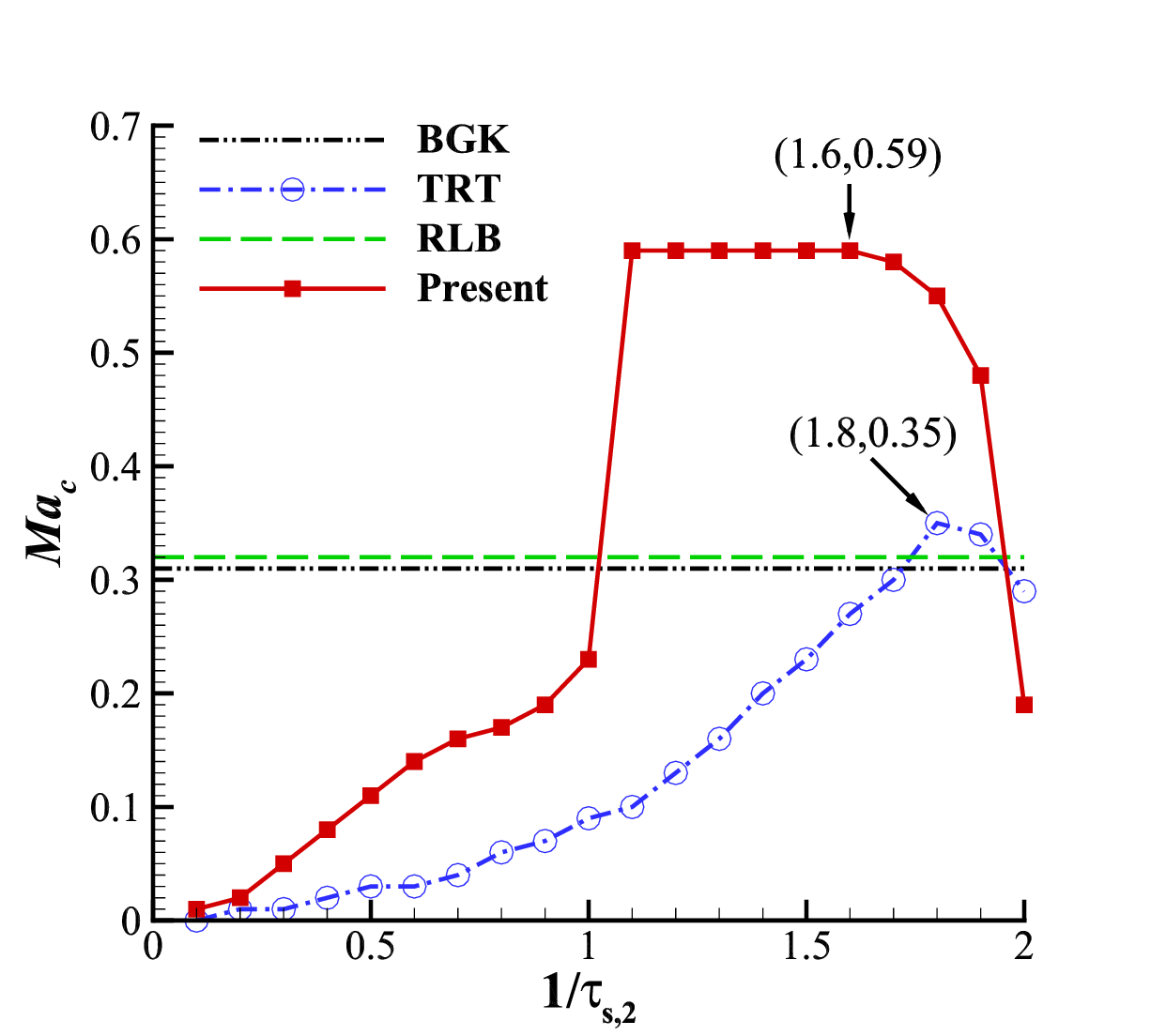}
		\caption{$Re=1\times 10^4$}
		\label{fig:sub2}
	\end{subfigure}
	\hfill
	\begin{subfigure}{0.32\textwidth}
		\includegraphics[width=\textwidth]{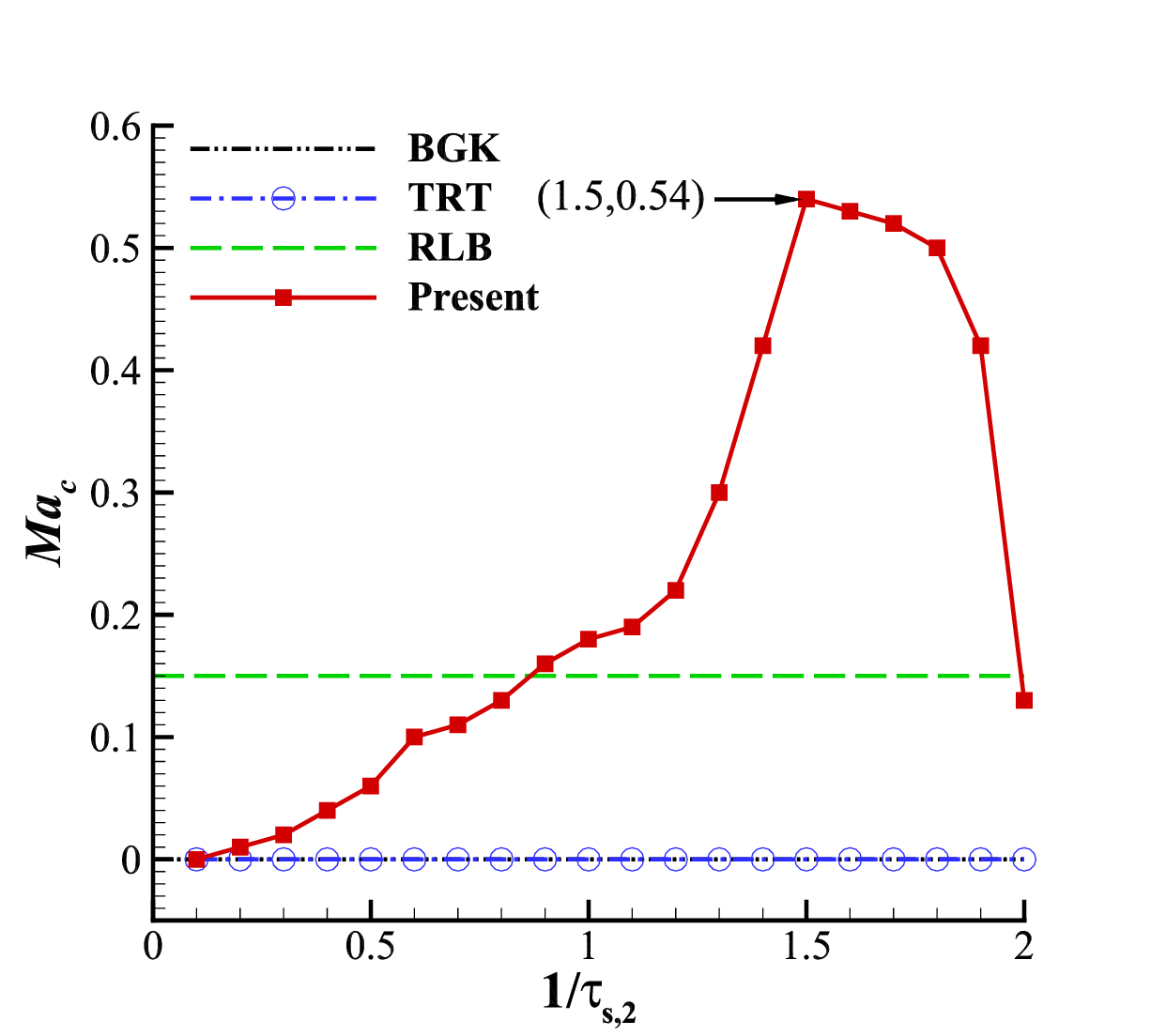}
		\caption{$Re=5\times 10^4$}
		\label{fig:sub3}
	\end{subfigure}
	
	\vspace{0.5cm}
	
	\begin{subfigure}{0.32\textwidth}
		\includegraphics[width=\textwidth]{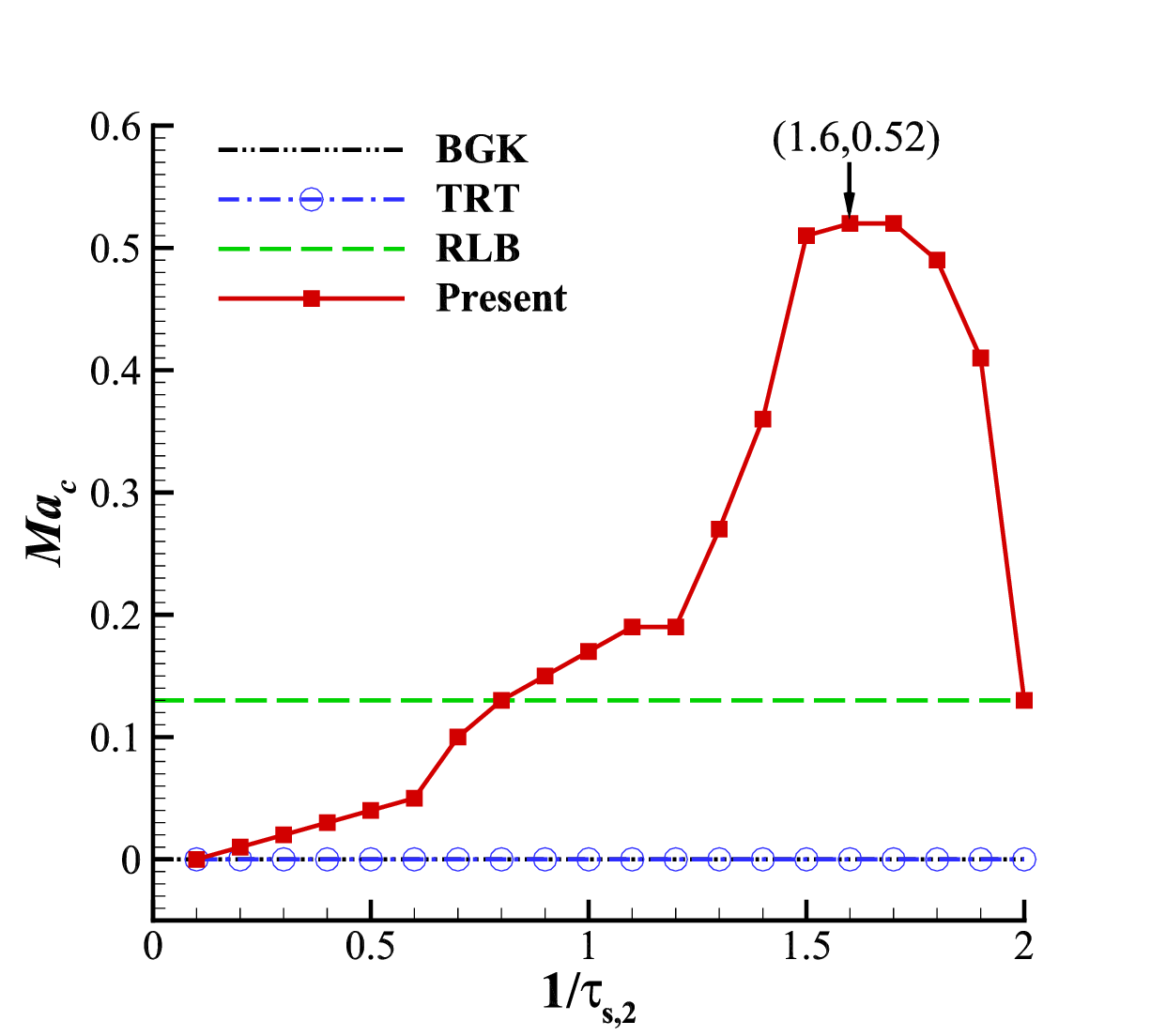}
		\caption{$Re=1\times 10^5$}
		\label{fig:sub4}
	\end{subfigure}
	\hfill
	\begin{subfigure}{0.32\textwidth}
		\includegraphics[width=\textwidth]{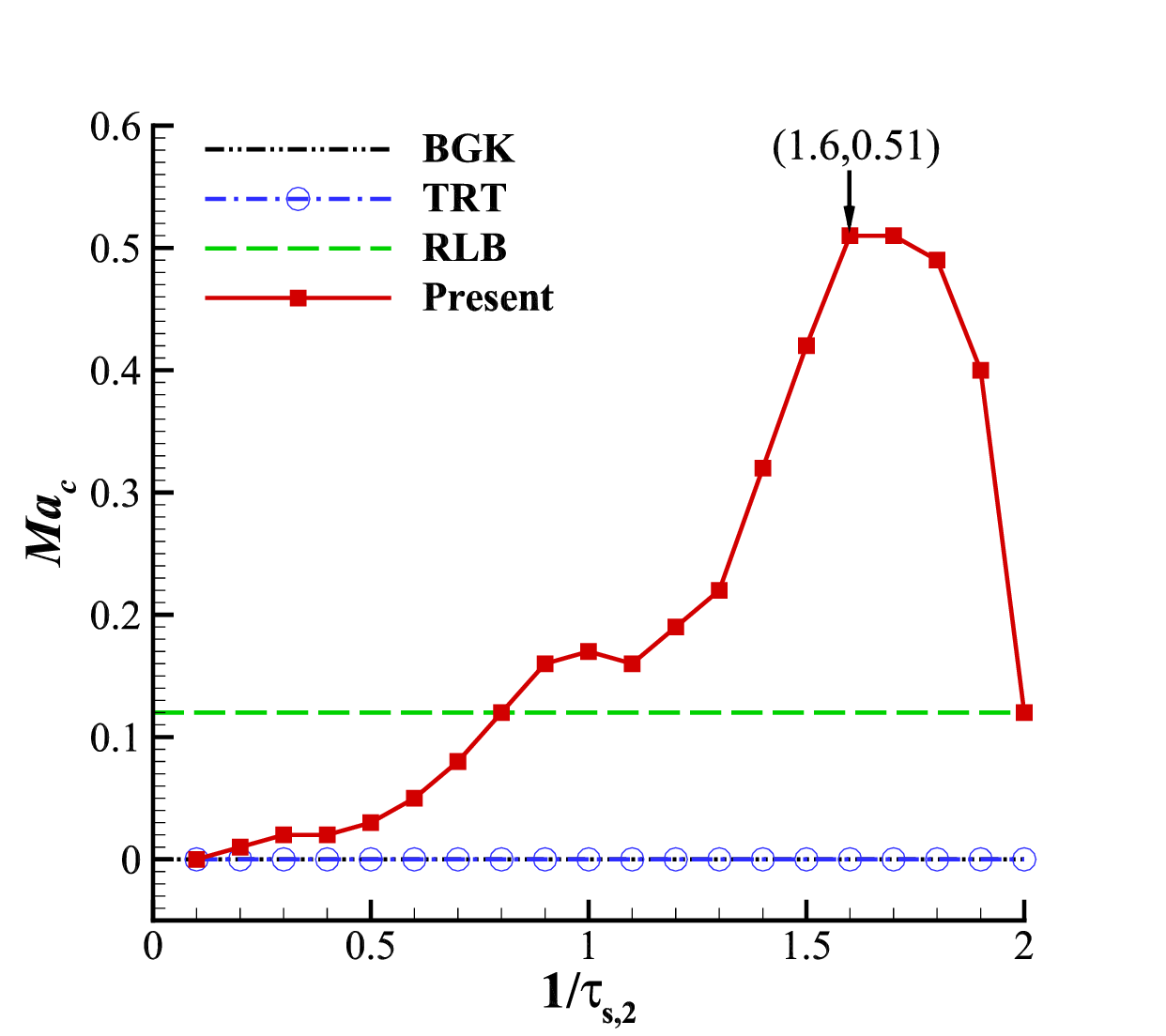}
		\caption{$Re=1\times 10^6$}
		\label{fig:sub5}
	\end{subfigure}
	\hfill
	\begin{subfigure}{0.32\textwidth}
		\includegraphics[width=\textwidth]{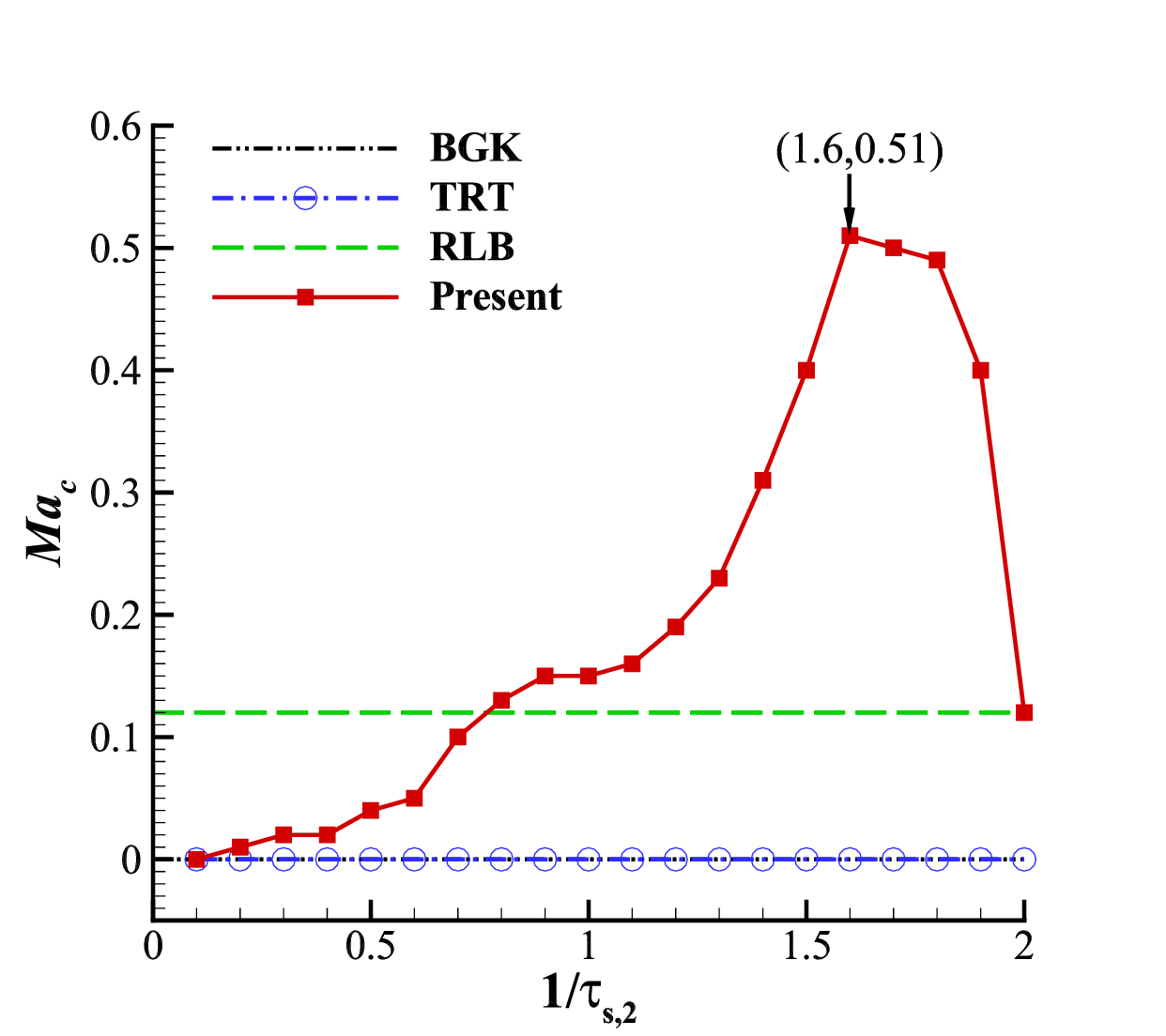}
		\caption{$Re=1\times 10^7$}
		\label{fig:sub6}
	\end{subfigure}
	\caption{Maximum critical Mach number $Ma_{c}$ versus different $1/\tau_{s,2}$ in the double shear layer simulations at various Reynolds numbers $Re$: (a) $Re=5\times 10^3$; (b) $Re=1\times 10^4$; (c) $Re=5\times 10^4$; (d) $Re=1\times 10^5$; (e) $Re=1\times 10^6$; (f) $Re=1\times 10^7$. The level of accuracy of $Ma_{c}$ is $\Delta Ma=0.01$. The horizontal lines without symbols indicate that the $Ma_{c}$ of these models (BGK and RLB) remain constant and unaffected by $\tau_{s,2}$. When the simulation result of a certain model fail to converge, $Ma_{c}=0$ is depicted. The specific values corresponding to several representative maximum $Ma_{c}$ are also explicitly indicated in the form of coordinates.}
	\label{test1_fig1}
\end{figure}

The comparison of stability among different models is illustrated in \hyperref[test1_fig1]{Figure~\ref{test1_fig1}}. The SRT-type models, i.e. the BGK and RLB models, are represented by lines without markers in the figure, indicating that they are not influenced by $\tau_{s,2}$. When $Re\geq 5\times 10^4$, both TRT and BGK models exhibit divergence, leading to a $Ma_{c}$ of zero. It can be observed that the present TRT-RLB model consistently demonstrates superior numerical stability compared to TRT. Notably, when $1/\tau_{s,2}$ ranges from 0.9 to 1.9 approximately, the stability of the TRT-RLB model exhibits a significant advantage over the RLB model. The stability of the present model reaches its optimal state around $1/\tau_{s,2} \approx 1.6$. At a Reynolds number of $Re=5\times10^3$, the maximum critical $Ma_{c}$ with the TRT-RLB model is 0.62. Even for larger Reynolds numbers ($5\times10^4 \sim 1\times10^7$), its maximum critical $Ma_{c}$ remains above $0.51$.
\subsection{2D decaying Taylor-Green vortex flow}

\quad To analyze the grid convergence and accuracy of the present model, we study the two-dimensional (2D) decaying Taylor-Green vortex flow. It is a well-known problem with a fully periodic flow domain, which is very suitable for the evaluation of collision models~\cite{strzelczyk2023study}. In this study, the computational domain is fixed in a 2D $\left[0,1\right]\times\left[0,1\right]$ region with periodic boundary conditions. The velocity and density fields of the flow is given as~\cite{pearson1965computational,kruger2017lattice}:
\begin{subequations}\label{eq_TG}
		\begin{align}
			u_x\left(\mathbf{x},t\right)=&-u_0 \cos\left(2\pi x\right)\sin\left(2\pi y\right)\exp^{-t/t_d},
		\\
			u_y\left(\mathbf{x},t\right)=&u_0 \cos\left(2\pi y\right)\sin\left(2\pi x\right)\exp^{-t/t_d},
		\\
			\rho\left(\mathbf{x},t\right)=&\rho_0-\rho_0\frac{u_0^2}{4c_s^2}\left[\cos\left(4\pi x\right)+\cos\left(4\pi y\right)\right]\exp^{-2t/t_d},
	\end{align}
\end{subequations}
where $u_0$ and $\rho_0$ are the initial velocity scale and density, respectively, and the vortex decay time $t_d=1/\left(8\nu\pi^2\right)$.

\begin{figure}[h]
	\centering
	\begin{subfigure}[b]{0.48\textwidth}
		\includegraphics[width=\textwidth]{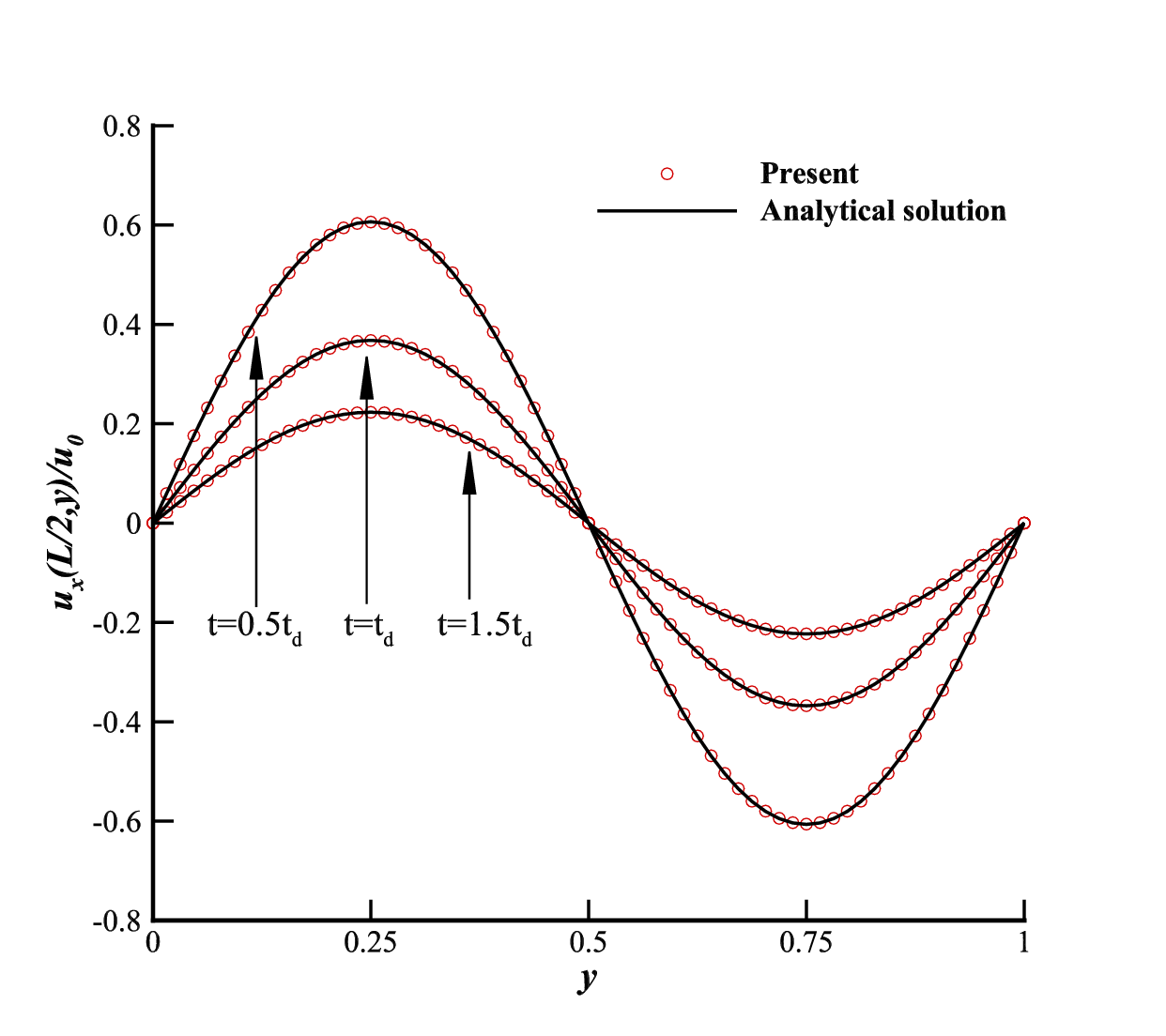}
	\end{subfigure}
	\hfill
	\begin{subfigure}[b]{0.48\textwidth}
		\includegraphics[width=\textwidth]{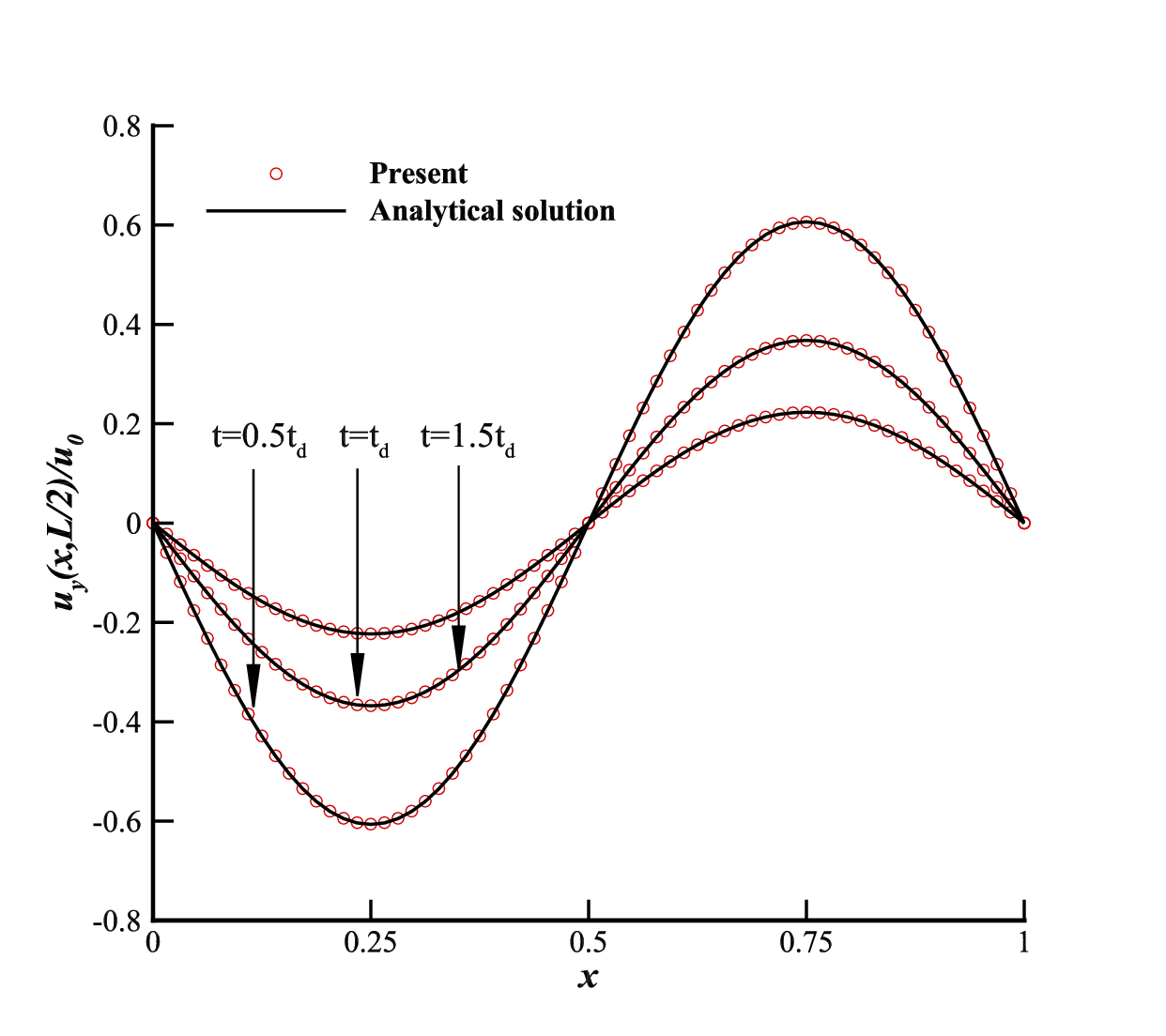}
	\end{subfigure}
	\caption{Comparisons between the results of $u_x$ (left) and $u_y$ (right) simulated by the present model and the analytical solutions for the 2D decaying Taylor-Green vortex flow on a $64\times64$ grid with $\nu=0.01$ at different time $t$.}
	\label{test2_fig1}
\end{figure}

In simulations, we set $u_c=u_0=0.01$, $\rho_0=1$, $\nu=0.01$ and $\Delta x^2/\Delta t=0.01\pi^2$ with the spatial step $\Delta x=L/N$, where $L=1$ is the domain side length and $N$ is the number of grids in the length of the domain. The magic parameter is set as $\Lambda_s=1/4$ to enhanced stability. To test the accuracy of the present model, the simulated velocities $u_x\left(L/2,y\right)$ and $u_y(x,L/2)$ at different time $t$ are compared with the analytical solution, Eq.~(\ref{eq_TG}), as presented in \hyperref[test2_fig1]{Figure~\ref{test2_fig1}}. The lattice size is set as $N=64$ here. As we can see, the numerical results match well with the analytical solutions. 

\begin{figure}[h]
	\centering
	\includegraphics[width=0.5\textwidth]{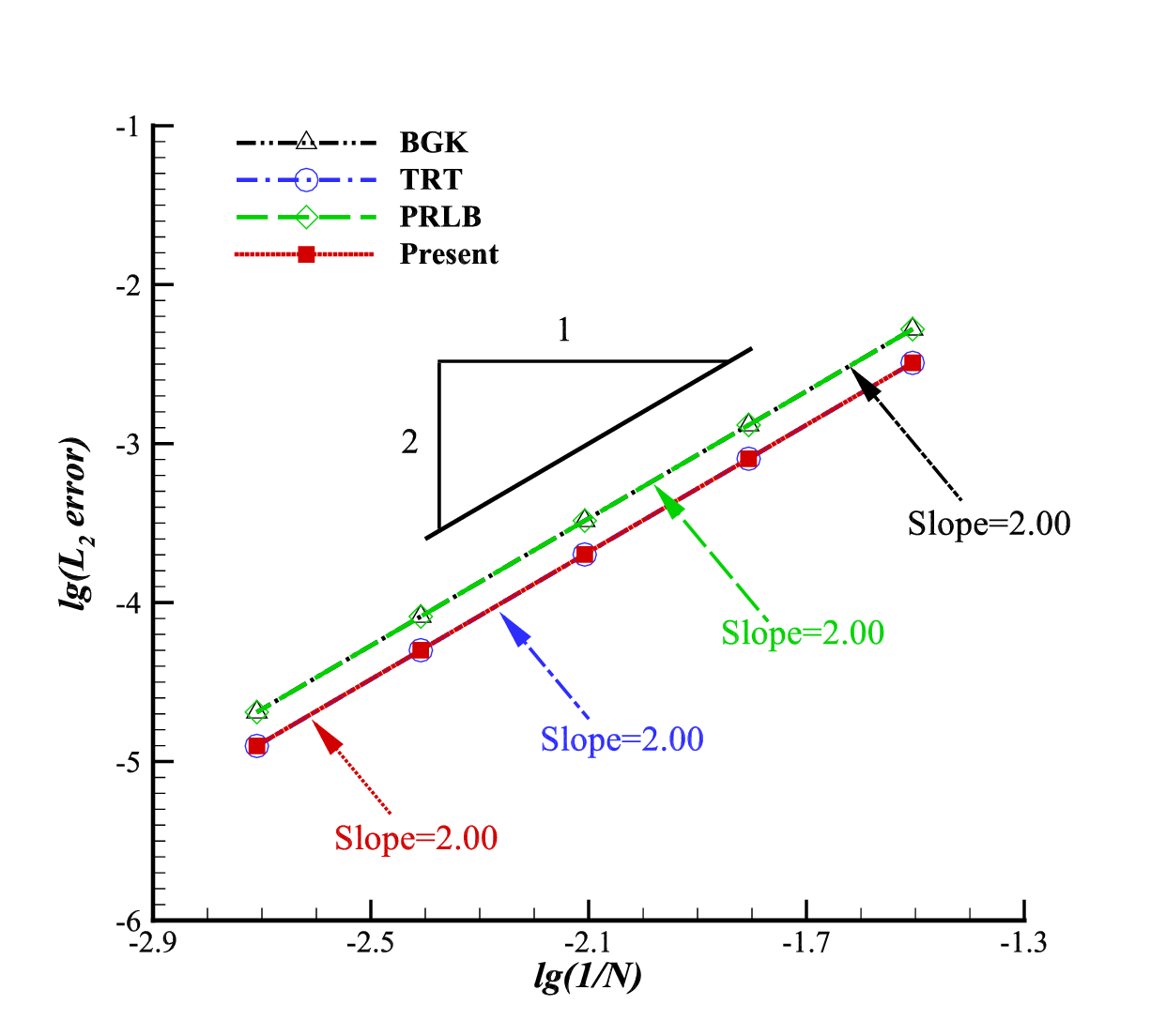}
	\caption{The $L_2$ error of the BGK, TRT, RLB and the present TRT-RLB models for the 2D decaying Taylor-Green vortex flow with $\nu=0.01$ at $t=t_d$.}
	\label{test2_fig2}
\end{figure}

Furthermore, the grid convergence orders of several different models, including BGK, TRT, RLB and the present model are investigated. Here, multiple grid sizes $N^2=$$32^2$, $64^2$, $128^2$, $256^2$, $512^2$ were taken into account.   These results are presented in \hyperref[test2_fig2]{Figure~\ref{test2_fig2}}. The findings indicate that the mentioned four models all possess spatial second-order convergence, and the present model possesses lower errors compared to BGK and RLB models.

\subsection{Force-driven Poiseuille flow}
\quad In Sec.~\ref{sec_CE}, the theoretical analysis indicates that the numerical slip of the HWBB scheme can be effectively eliminated by the present TRT-RLB model with a specific free relaxation parameter. To validate the expression Eq.~(\ref{eq_a29}), here we conduct the simulations of the force-driven Poiseuille flow, which can be described by Eqs.~(\ref{eq_a1}) and (\ref{eq_a3}).

In simulations, we configure the relevant parameters as follows: the channel width $L=1$, the particle speed $c=1$, the characteristic velocity $u_c=0.1$, the initial density $\rho_0=1$, and the Reynolds number $Re=1$. The grid nodes are setting in a $N_x\times N_y=4\times32$ domain, and the spatial step $\Delta x=L/N_y$.

\begin{figure}[h]
	\centering
	\includegraphics[width=0.5\textwidth]{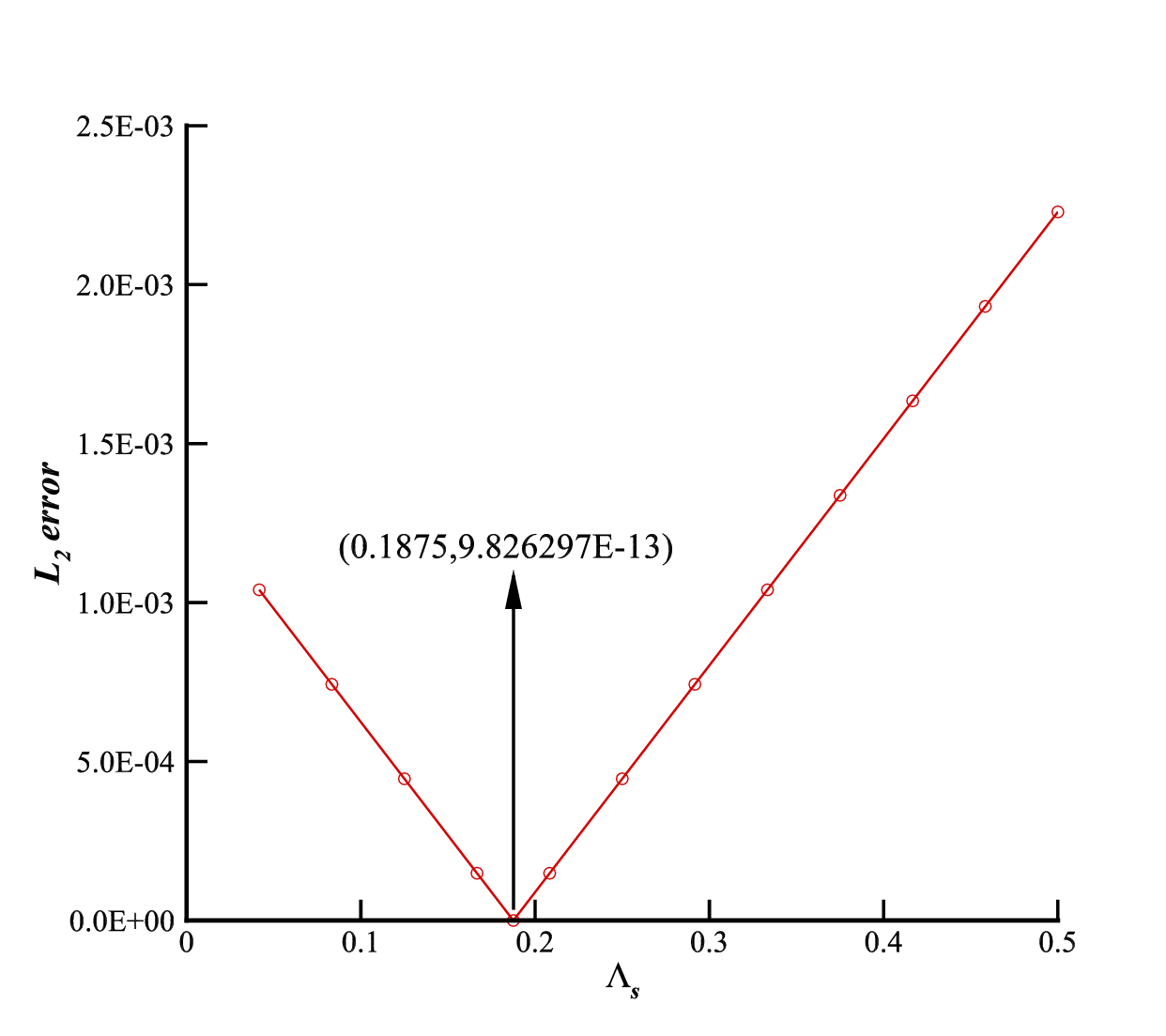}
	\caption{The $L_2$ error at different magic parameter $\Lambda_s$ simulated by the proposed TRT-RLB model for the force-driven Poiseuille flow.}
	\label{test3_fig1}
\end{figure}

\begin{figure}[h]
	\centering
	\includegraphics[width=0.5\textwidth]{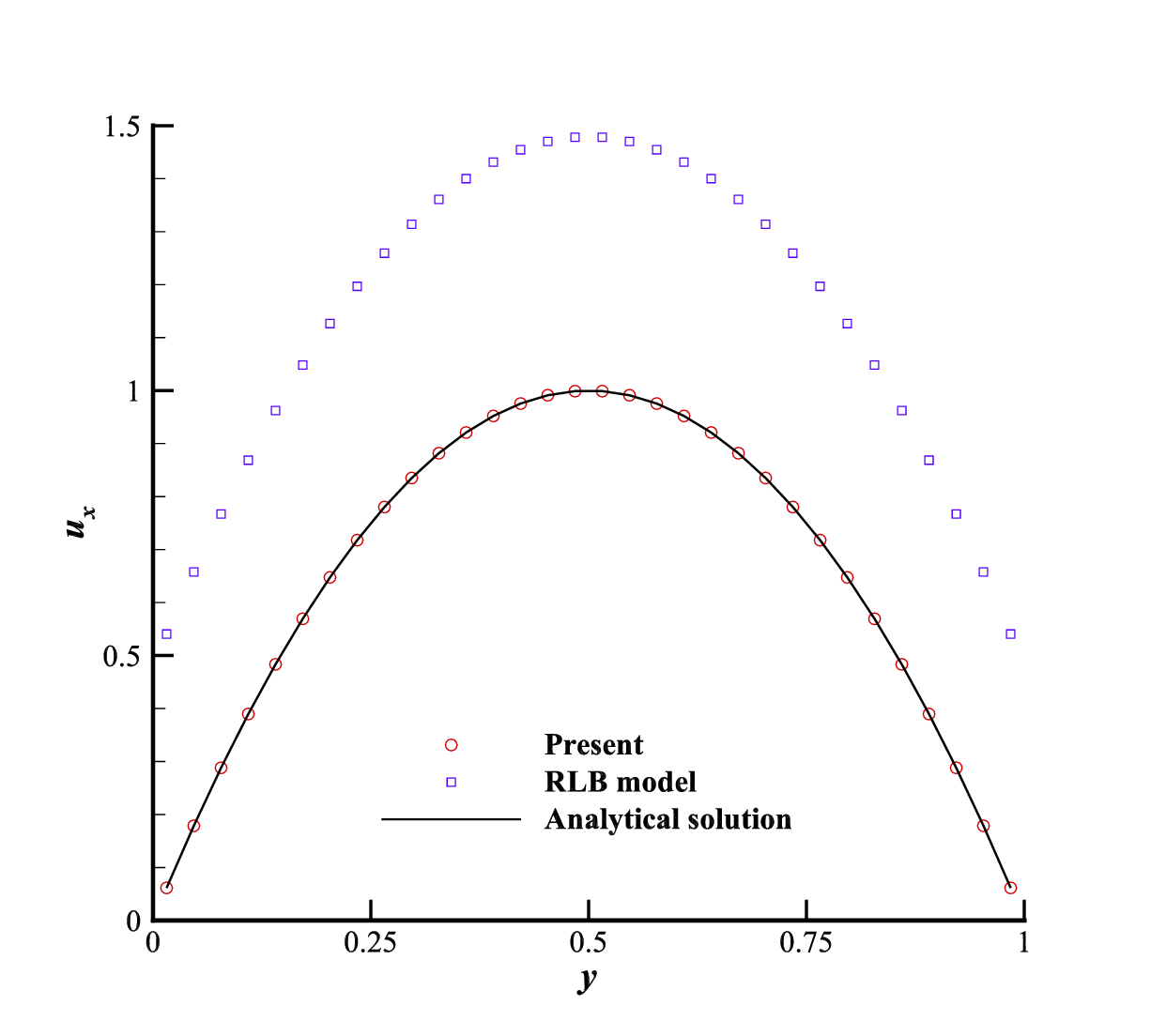}
	\caption{Comparisons of the horizontal velocity profiles between the analytical solution and the simulation results obtained by the present model and the RLB model.}
	\label{test3_fig2}
\end{figure}

The analysis of the simulation results primarily consists of two parts. Firstly, the impacts of the magic parameter $\Lambda_s$ in the accuracy of present model are investigated and results are shown in \hyperref[test3_fig1]{Figure~\ref{test3_fig1}}. As observed in this figure, the optimal magic parameter that achieves complete elimination ($L_2\sim O(10^{-13})$) of numerical slip is 0.1875, consistent with the theoretical analysis presented in Sec.~\ref{sec_CE}. Furthermore, the RLB model are also used to simulated the present problem. The theoretical solution of the horizontal velocity profile is compared with the simulation results of the RLB model~\cite{coreixas2017recursive} and the proposed TRT-RLB model with $\Lambda_s=3/16$, as depicted in \hyperref[test3_fig2]{Figure~\ref{test3_fig2}}. The results indicate that the numerical slip elimination capability of the proposed TRT-RLB model does indeed not exist in the RLB model.

\subsection{Creeping flow past a square cylinder}
\begin{figure}[h]
	\centering
	\includegraphics[width=0.5\textwidth]{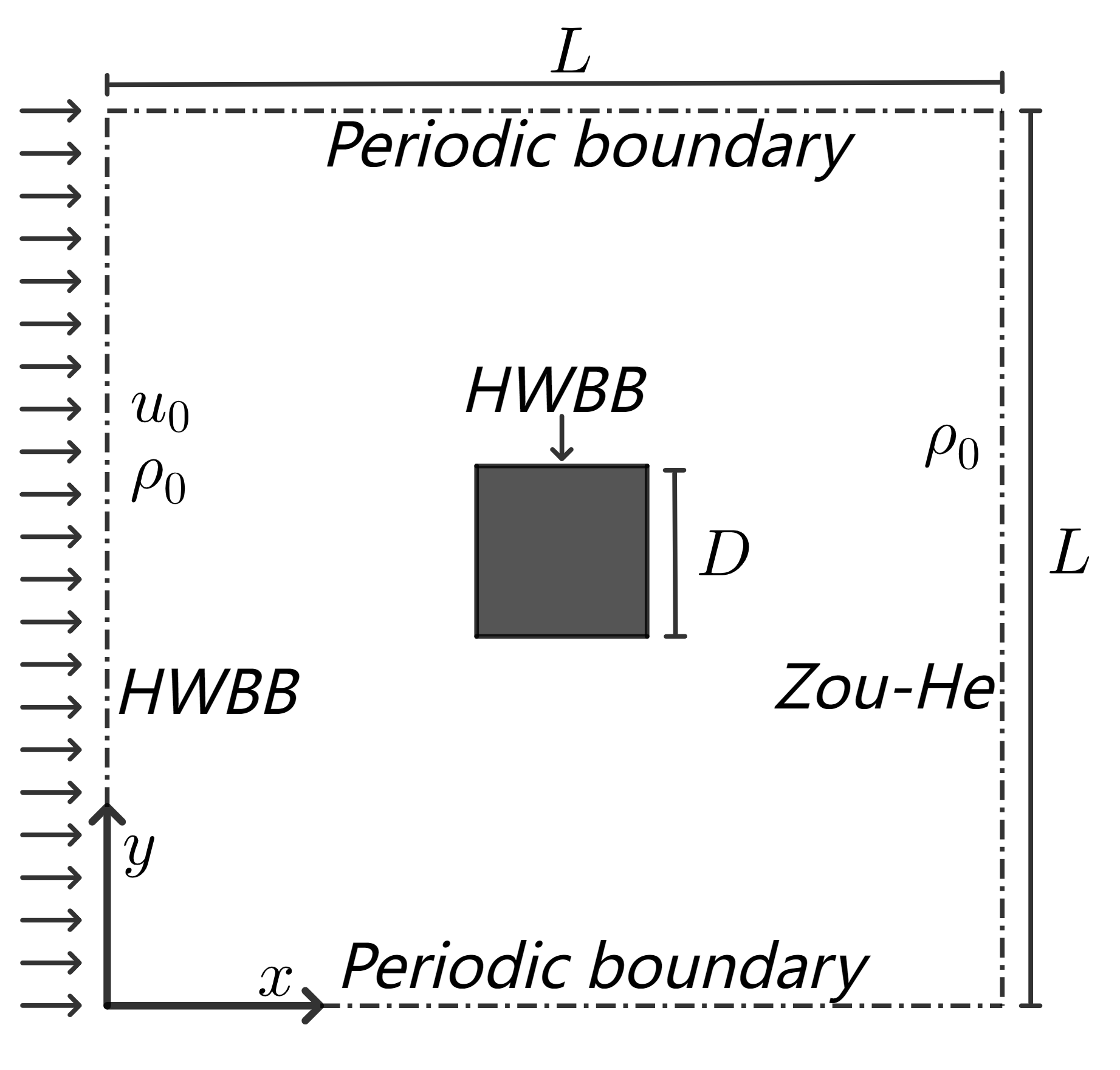}
	\caption{Schematic diagram of the creeping flow past a square cylinder.}
	\label{test4_fig1}
\end{figure}

\quad In this section, we will exemplify the advantages of the current TRT-RLB model in simulating low Reynolds number flows by conducting a numerical experiment on creeping flow past a square cylinder. \hyperref[test4_fig1]{Figure~\ref{test4_fig1}} illustrates the computational domain geometry, coordinate system, and boundary conditions. The simulation is conducted within a square region with side length $L$. The inflow enters the computational domain from the left side with a uniform velocity $u_c$, then passes over and through a square cylinder with side length $D$. The center of the square cylinder exactly coincides with the center of the computational domain. The top and bottom boundaries of the computational region are set as periodic boundary conditions. The left boundary employs the HWBB scheme to implement a constant velocity boundary condition. The right boundary utilizes the Zou and He method to implement a pressure-constant outlet boundary condition. At the cylinder surface, the no-slip boundary condition is enforced using the HWBB scheme. In the numerical simulation of the present problem, three dimensionless parameters, namely the Reynolds number ($Re$), Mach number ($Ma$), and the incompressibility viscosity coefficient ($\mathcal{T}$), significantly influence the numerical results. The incompressibility viscosity parameter, introduced by Gsell et~al.~\cite{gsell2021lattice}, is expressed as follows:
\begin{equation}
	\mathcal{T}=\frac{Ma^2}{Re}.
\end{equation}
In this section, the termination criterion for the simulation runs of all cases is set as $L_2< 10^{-9}$, to accommodate the smaller characteristic velocities encountered in creeping flow conditions. In simulations, the initial density $\rho_0=1$, the lattice speed $c=\Delta x/\Delta t=1$ with $\Delta x=1$, the square side length $L=50D$ and the magic parameter is set as $\Lambda_s=3/16$ to eliminate the numerical slip as discussed in the previous section.

\begin{figure}[h]
	\centering
	\begin{subfigure}[b]{0.48\textwidth}
		\includegraphics[width=\textwidth]{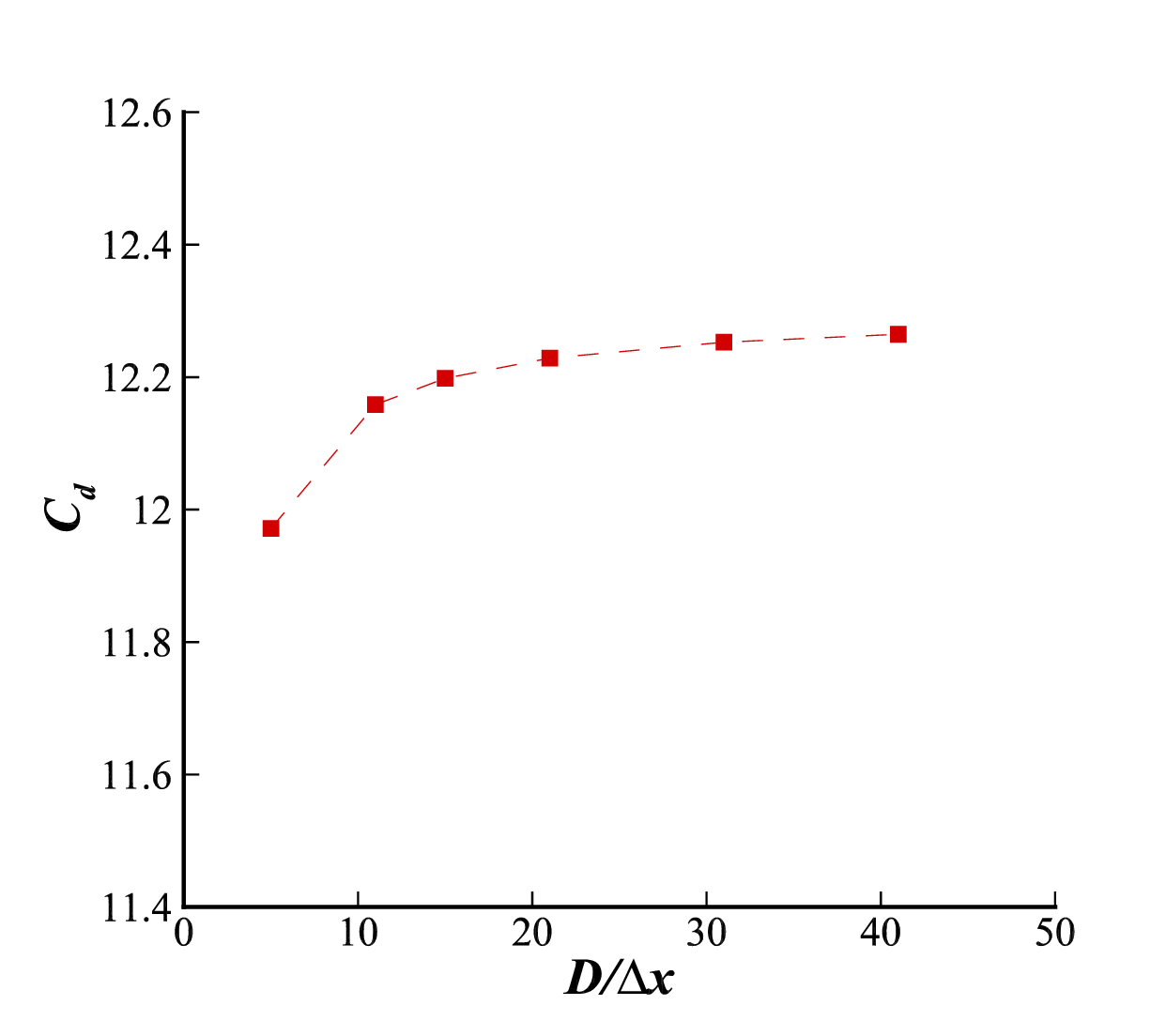}
		\caption{}
	\end{subfigure}
	\hfill
	\begin{subfigure}[b]{0.48\textwidth}
		\includegraphics[width=\textwidth]{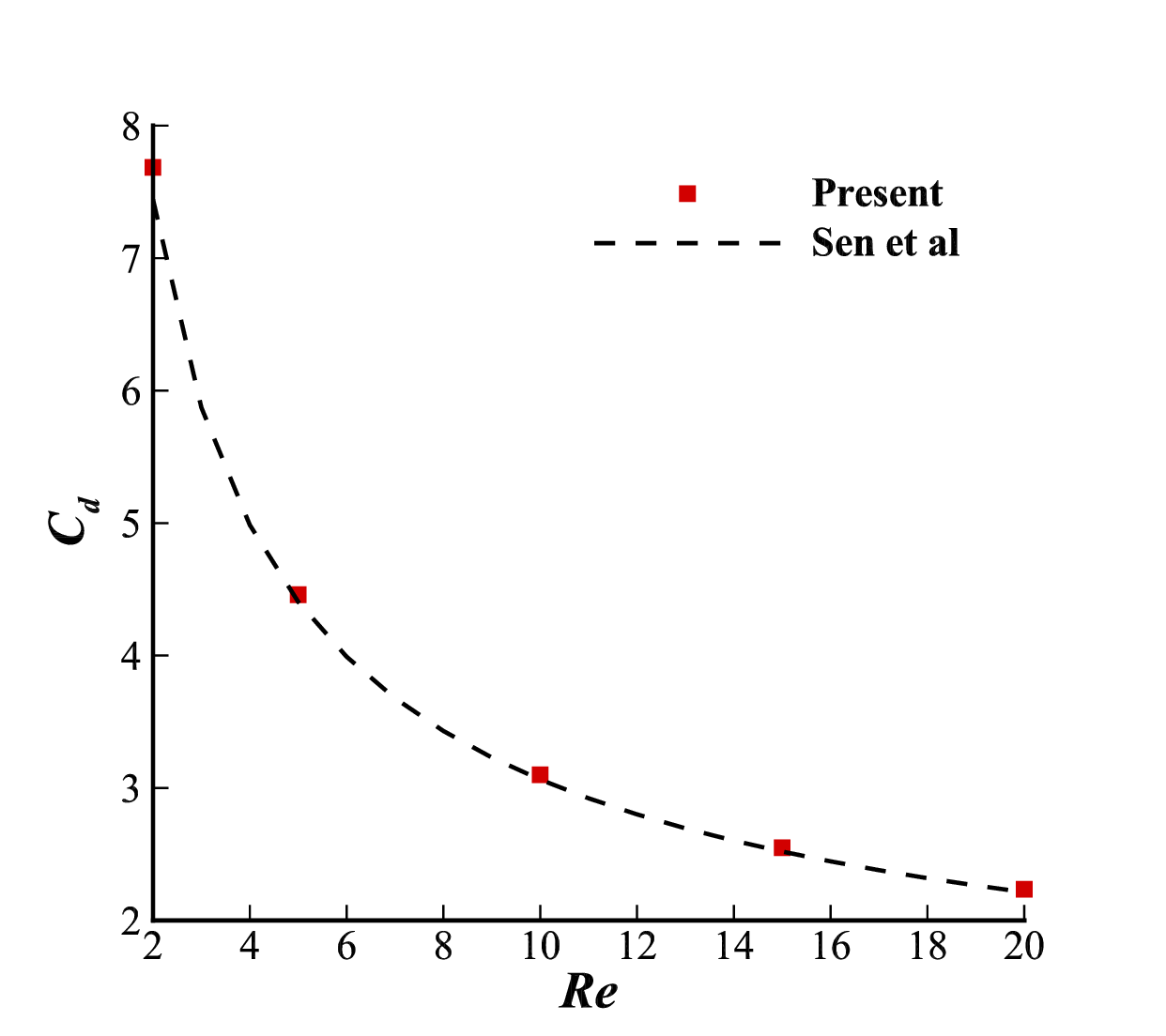}
		\caption{}
	\end{subfigure}
	\caption{The flow around a square cylinder problem simulated by the TRT-RLB model: (a) relationship between the drag coefficient $C_d$ and grid resolution $D/\Delta x$ with Re=1 and Ma=0.1, and (b) results of the drag coefficient $C_d$ at various Reynolds numbers $Re$ with $D/\Delta x=21$.}
	\label{test4_fig2}
\end{figure}

The simulation and discussion comprise three parts. Firstly, a grid independence test was conducted. Here, we set the Reynolds number $Re=1$, and the Mach number $Ma=0.1$. Six different cases with varying grid refinement parameter $D/\Delta x$ were tested, and the corresponding drag coefficients $C_d$ were recorded, where the drag coefficients $C_d$ is defined as
\begin{equation}
	C_d=\frac{2F_x}{\rho_0 u_0^2 D},
\end{equation}
where
\begin{equation}
	F_x=\frac{\Delta x^2}{\Delta t}\sum_{i\in I_w}e_{ix}\left[f_{\overline{i}}^{*}\left(x_s,t\right)+f_{i}^{*}\left(x_f,t\right)\right],
\end{equation}
with $I_w$ representing the collection of directions on the outermost fluid nodes $x_f$ pointing towards the wall, $x_{s}$ being the solid nodes that the post-collision $f_{i}^{*}\left(x_f,t\right)$ can arrive, and $\overline{i}$ denoting the opposite direction to $i$. The results are presented in \hyperref[test4_fig2]{Fig~\ref{test4_fig2}(a)}. It can be observed that the drag coefficient $C_d$ gradually stabilizes as the grid refinement parameter $D/\Delta x$ increases. In order to control computational costs while satisfying accuracy requirements, this paper adopts a grid size where $D=21\Delta x$ for subsequent simulations.

Next, the accuracy of the model was validated. We simulated a set of five cases within the Reynolds number range of 2 to 20 and compared them with the empirical drag law proposed by Sen et~al.~\cite{sen2011flow}, which was based on high-precision finite element method simulation results. This law can be expressed by the following formula:
\begin{equation}
	C_d=0.7496+10.5767Re^{-0.66}, \quad Re \in [2,40].
\end{equation}
The comparative results are presented in \hyperref[test4_fig2]{Fig~\ref{test4_fig2}(b)}. It is evident from these results that our model demonstrates accuracy in simulating flow around a square cylinder.

Furthermore, the performance of the model in low Reynolds number scenarios ($Re\le0.1$) was investigated. According to the research conclusions of Gsell et~al.~\cite{gsell2021lattice}, under such Reynolds number settings, maintaining the incompressible viscosity parameter $\mathcal{T}\ll1$ is sufficient to ensure program stability. Therefore, we set $\mathcal{T}=0.1$ in this context. The settings for other parameters remain consistent with previous configurations. Upon achieving program stability, the Reynolds number-independent viscous drag coefficient $C_{d,\mu}$ will be recorded, where the so-called viscous drag coefficient $C_{d,\mu}$ is defined as 
\begin{equation}
	C_{d,\mu}={C_d}{Re},
\end{equation}
which is particularly convenient in viscous stress-dominated low Reynolds number flows. In addition to the TRT-RLB model, the RLB model has also been employed for simulations. Results are all presented in \hyperref[tab1]{Table~\ref{tab1}}. It is observed that as the Reynolds number decreases from the order of $10^{-1}$ to $10^{-7}$, the viscous drag coefficients $C_{d,\mu}$ of the TRT-RLB simulations rapidly stabilizes and shows no significant change at $Re\leq 10^{-3}$. This outcome is reasonable and validates the accuracy and stability of the proposed model at ultra-low Reynolds numbers. However, the simulation results of the RLB model obviously deviate from the stable solution of $8.252$ provided by the TRT-RLB model at $Re=0.1$, and the program diverges at $Re\leq 10^{-3}$.  Furthermore, it can be observed that when employing the TRT-RLB model, the value of $\tau_{s,1}$ can be set as high as $36374$, while in SRT-type models like BGK and RLB, it is a prerequisite that $\tau_{s,1}<3$ as analyzed by Gsell et~al.~\cite{gsell2021lattice}.
\begin{table}[h]
	\centering
	\caption{The simulation results for the flow around a square cylinder at low Reynolds number, with $\mathcal{T}=0.1$, $D/\Delta x=21$ and $L=50D$. (The symbol ``-'' indicates divergence in simulation.)}
	\label{tab1}%
	\begin{ruledtabular}
		\begin{tabular}{ccccccc}
			\multicolumn{1}{c}{\multirow{2}{*}{$Re$}} & \multicolumn{1}{c}{\multirow{2}{*}{$Ma$}} & \multicolumn{1}{c}{\multirow{2}{*}{$Kn$}} & \multicolumn{1}{c}{\multirow{2}{*}{$\tau_{s,1}$}} & \multicolumn{2}{c}{$C_{d,\mu}$} & \multirow{2}{*}{} \\
			\cline{5-6} & & & & TRT-RLB & RLB & \\
			$10^{-1}$ & $1.00\times10^{-1}$ & $1$ & $37$ & 8.449 & 4.275 & \\
			$10^{-2}$ & $3.16\times10^{-2}$ & $3.16$ & $116$ & 8.260 & 1.610 & \\
			$10^{-3}$ & $1.00\times10^{-2}$ & $10$ & $364$ & 8.253 & - & \\
			$10^{-4}$ & $3.16\times10^{-3}$ & $31.6$ & $1151$ & 8.253 & - & \\
			$10^{-5}$ & $1.00\times10^{-3}$ & $100$ & $3638$ & 8.252 & - & \\
			$10^{-6}$ & $3.16\times10^{-4}$ & $316$ & $11503$ & 8.252 & - & \\
			$10^{-7}$ & $1.00\times10^{-4}$ & $1000$ & $36374$ & 8.252 & - & \\
		\end{tabular}
	\end{ruledtabular}
\end{table}
\section{Conclusion}\label{section4}
\quad In this paper, we proposed a two-relaxation-time regularized lattice Boltzmann (TRT-RLB) model for simulating the weakly compressible isothermal flow. In our model, the non-equilibrium distribution function is expanded via Hermite polynomials up to the third order. The first and second order non-equilibrium relax with a viscosity-related relaxation time, $\tau_{s,1}$, while the third order relaxes with a free relaxation time, $\tau_{s,2}$. Additionally, two enhancements were concurrently implemented to eliminate the cubic velocity error term inherent in the standard lattice. Specifically, the equilibrium distribution function was expanded to the third order, and a compensatory term, denoted as $G_i$, was introduced. Through a CE analysis, it has been demonstrated that this model can accurately recover the macroscopic Navier-Stokes equations. Furthermore, it was found that the parameter $\tau_{s,2}$ exerts no influence on the recovery process, indicating its role as a free parameter.

Previous research~\cite{gsell2021lattice,ginzburg2008study} has indicated that single relaxation models, exemplified by the BGK model, exhibit viscosity-dependent numerical slip due to discretization errors. The TRT model addresses these errors by adjusting a magic parameter, which can eliminate numerical slip from the HWBB scheme when fixed at 3/16. This paper, through theoretical analysis, proves that our proposed TRT-RLB model possesses the same advantages as the TRT model.

Moreover, our model's numerical stability surpasses that of the BGK, TRT, and RLB models, as evidenced by simulations of a double shear layer problem with periodic boundary conditions. At a Reynolds numbers $Re=5\times 10^{3}$, the TRT-RLB model maintains stability up to a maximum critical Mach number $Ma_c=0.62$, compared to just 0.38 and 0.49 for the TRT and RLB models, respectively. The superiority of the TRT-RLB model becomes more pronounced at high Reynolds numbers ($Re=$$5\times 10^4$ to $10^7$). The maximum $Ma_c$ remains consistently above 0.51 for the proposed TRT-RLB model, while both the BGK and TRT models have diverged. In comparison, the maximum $Ma_c$ for the RLB model is only about $0.12$ to $0.15$ at high Reynolds numbers.

Additionally, the model's second-order spatial accuracy is validated through simulations of 2D decaying Taylor-Green vortex flow. The capability of the TRT-RLB model to eliminate numerical slip present in the HWBB scheme is corroborated through simulations of force-driven Poiseuille flow.

Lastly, the robustness of the proposed model at ultra-low Reynolds numbers is affirmed through the simulations of creeping flow around a square cylinder. Our model is capable of accurately and reliably handling the problems with Reynolds numbers as low as $10^{-7}$. In contrast, the RLB model converges only when $Re > 10^{-2}$, the steady-state solution of the RLB model significantly deviates from the convergent value observed in the TRT-RLB model when $Re=0.1$, and the value of the steady-state solution exhibits no convergence trend with the increase in Reynolds number. 

The article concludes that the TRT-RLB model proposed herein exhibits significant robustness and adaptability, suggesting its potential to become a highly reliable tool in the field of computational fluid dynamics. Particularly, its performance in complex fluid domains, such as non-Newtonian fluid mechanics, is anticipated with keen interest. This model's resilience and versatility make it a promising candidate for various applications, especially in addressing the challenges posed by complex fluids.

%%%% Acknowledgments %%%%%%%%

\begin{acknowledgments}
This work is financially supported by the National Natural Science Foundation of China (Grant
Nos. 12101527, 12271464 and 11971414), the Science and Technology Innovation Program of Hunan Province (Program No. 2021RC2096), Project of Scientific Research Fund of Hunan Provincial Science and Technology Department (Grant No. 21B0159) and the Natural Science Foundation for Distinguished Young Scholars of Hunan Province (Grant No. 2023JJ10038). 
\end{acknowledgments}

%\section*{Data Availability Statement}

\appendix

%\section{Appendixes}

\nocite{*}
\bibliography{aipsamp}% Produces the bibliography via BibTeX.

\end{document}